 \documentclass[twocolumn]{aastex62}
\usepackage{enumitem}
\usepackage{hhline}
\shorttitle{Synthetic Spectra of 3D Pair--Instability Supernovae}
\shortauthors{Chatzopoulos, Gilmer, Wollaeger, Fr\"ohlich, \& Even}

\begin{document}

\title{SYNTHETIC SPECTRA OF PAIR--INSTABILITY SUPERNOVAE IN 3D}

\correspondingauthor{Emmanouil Chatzopoulos}
\email{chatzopoulos@phys.lsu.edu}

\author[0000-0002-0786-7307]{E. Chatzopoulos}
\affiliation{Department of Physics \& Astronomy, Louisiana State University, Baton Rouge, LA, 70803, USA}
\affiliation{Hearne Institute of Theoretical Physics, Louisiana State University, Baton Rouge, LA, 70803, USA}

\author[0000-0001-5396-6771]{Matthew S. Gilmer}
\affiliation{Center for Theoretical Astrophysics, Los Alamos National Laboratory, Los Alamos, NM 87544, USA}

\author{Ryan T. Wollaeger}
\affiliation{Center for Theoretical Astrophysics, Los Alamos National Laboratory, Los Alamos, NM 87544, USA}

\author[0000-0003-0191-2477]{Carla Fr\"ohlich}
\affiliation{Department of Physics, North Carolina State University, Raleigh, NC 27695, USA}

\author{Wesley P. Even}
\affiliation{Center for Theoretical Astrophysics, Los Alamos National Laboratory, Los Alamos, NM 87544, USA}



\begin{abstract}

Pair--Instability Supernovae (PISNe) may signal the deaths of extremely massive stars in the local Universe or massive primordial stars after the end 
of the Cosmic Dark Ages. Hydrodynamic simulations of these explosions, performed in 1D, 2D, and 3D geometry, have revealed 
the strong dependence of mixing in the PISN ejecta on dimensionality. This chemical rearrangement is mainly driven by 
Rayleigh-–Taylor instabilities that start to grow shortly after the collapse of the carbon--oxygen core. We investigate the effects of such 
mixing on the spectroscopic evolution of PISNe by post--processing explosion profiles with the radiation diffusion--equilibrium code 
{\it SNEC} and the implicit Monte Carlo -- discrete diffusion Monte Carlo (IMC--DDMC) radiation transport code {\it SuperNu}. 
The first 3D radiation transport calculation of a PISN explosion is presented yielding viewing angle--dependent synthetic spectra and lightcurves.
We find that while 2D and 3D mixing does not significantly affect the lightcurves of PISNe, their spectroscopic and color evolution is impacted. Strong features
of intermediate mass elements dominated by silicon, magnesium and oxygen appear at different phases and reach different intensities depending
on the extent of mixing in the silicon/oxygen interface of the PISN ejecta. On the other hand, we do not find a significant dependence of 
PISN lightcurves and spectra on viewing angle. Our results showcase the capabilities of {\it SuperNu} to handle 3D radiation transport and highlight
the importance of modeling time--series of spectra in identifying PISNe with future missions.

\end{abstract}



\section{Introduction} \label{intro}

Pair--Instability Supernovae (PISNe; \citealt{1967PhRvL..18..379B,1967ApJ...148..803R,1983A&A...119...61O}) are thought
to mark the catastrophic explosions of very massive stars that form carbon--oxygen (CO) cores with masses $M_{\rm CO} >$~60~$M_{\odot}$.
The collapse of these massive CO cores is triggered by a softening of the equation of state (EoS) where the adiabatic index, $\gamma_{\rm ad}$,
falls below 4/3 due to rapid electron--positron (e$^{-}$--e$^{+}$) pair production. As a result, a large amount of carbon and oxygen fuel
is ignited leading to the production of large sums ($>$~1~$M_{\odot}$) of radioactive $^{56}$Ni. The decay of $^{56}$Ni can subsequently
heat the expanding supernova (SN) ejecta and may even lead to superluminous lightcurves (LCs) with peak luminosities 
$L_{\rm max} > 10^{44}$~erg~s$^{-1}$ \citep{2002ApJ...567..532H,2011ApJ...734..102K}.

The extreme luminosities that can be reached in PISNe suggest that these explosions may be related to some events in the superluminous
supernova (SLSN) class \citep{2012Sci...337..927G,2018arXiv181201428G,2018SSRv..214...59M}. For instance,
the hydrogen--poor (SLSN--I) SN~2007bi \citep{2009Natur.462..624G} and the slowly--evolving Type II SN OGLE14--073 \citep{2018MNRAS.479.3106K}
are often discussed as PISN candidates (see also \citealt{2017MNRAS.468.4642I,2017ApJ...835...13J}). 
All SLSNe observed to date, however, are found at host environments with metallicities $Z >$~0.1~$Z_{\odot}$
\citep{2011ApJ...727...15N,2014ApJ...787..138L} suggesting a very large zero--age main--sequence (ZAMS) mass is needed ($M_{\rm ZAMS} \gtrapprox$~250--300~$M_{\odot}$) 
to overcome the effects of radiatively--driven mass--loss and to form CO cores that are massive enough to be susceptible to the pair instability \citep{2007A&A...475L..19L}.

In addition, synthetic spectra of PISNe have difficulty in matching the observed spectra of SLSNe at contemporaneous epochs 
\citep{2013MNRAS.428.3227D,2015ApJ...799...18C,2019MNRAS.tmp..259M}. Some proposed ways to help mitigate these issues 
include enhanced mixing due to rapid progenitor rotation allowing PISNe to be encountered
at a considerably lower $M_{\rm ZAMS}$ \citep{2012ApJ...748...42C,2012A&A...542A.113Y} and large--scale outward mixing of 
$^{56}$Ni \citep{2015MNRAS.454.4357K} (see also \citealt{2014A&A...565A..70K,2014A&A...566A.146K}). 
A softer version of PISN that does not lead to the complete disruption of the progenitor star is the pulsational pair--instability supernova mechanism 
(PPISN; \citealt{2007Natur.450..390W}), encountered for a narrow range of $M_{\rm ZAMS}$ below the limit for full--fledged PISN. 
PPISN can result in the ejection of multiple shells that can interact with each other yielding several transient events from the same progenitor
and, occasionally, to very bright LCs akin to those of SLSNe \citep{2017ApJ...836..244W}. These theoretical implications, coupled
with observations of massive stars up to $\sim$~300~$M_{\odot}$ within the young star clusters NGC3603 and R136 in the Milky Way galaxy 
\citep{2010MNRAS.408..731C}, encourage the notion that while PISNe and PPISNe events must be very rare in the contemporary Universe, they
cannot be ruled out.

In addition, very massive ($>$~200~$M_{\odot}$) metal--poor stars in the early Universe are more likely PISN progenitors 
\citep{2007A&A...461..571H,2011ApJ...728..129J,2012MNRAS.422.2701P,2012MNRAS.422..290S}. Population III star formation
simulations suggest that the first generation of stars had a top--heavy initial mass function (IMF) \citep{1998ApJ...508..518A,2002ApJ...564...23B,2004ARA&A..42...79B} 
with a large fraction of them well within the mass limit to encounter PISN \citep{2008Sci...321..669Y,2009Natur.459...49B,2012MNRAS.422..290S}.
Given that some PISN models imply very bright lightcurves, these explosions could be detected at large redshifts with upcoming missions such as the 
{\it James Webb Space Telescope} ({\it JWST}) and {\it WFIRST} \citep{2005ApJ...633.1031S,2012ApJ...755...72H,2013ApJ...777..110W,2015ApJ...805...44S}.

The possible link between PISNe, SLSNe and the evolution of very massive stars has driven many efforts to study this mechanism in detail by making use
of numerical supercomputer simulations in 2D and 3D \citep{2014ApJ...792...28C,2014ApJ...792...44C}. In particular, the role
of mixing induced by Rayleigh-–Taylor (RT) instabilities in 3D and their impact on PISN LCs (\citealt{2017ApJ...846..100G}; hereafter G17), 
as well as the effects of rapid progenitor rotation on the energetics, dynamics and nucleosynthetic signatures of the explosion have been investigated 
\citep{2013ApJ...776..129C}. As mentioned before, radiation transfer calculations yielding synthetic spectra and LCs for different PISN progenitor properties 
have also been presented \citep{2013MNRAS.428.3227D,2015ApJ...799...18C,2019MNRAS.tmp..259M}. 

In this work we study the effects of multidimensional mixing on the spectroscopic properties of PISNe by calculating synthetic LCs and spectra
for progenitor profiles computed in 1D, 2D and 3D hydrodynamics simulations. To do so, we extract profiles corresponding to regions
of high inward and outward mixing of silicon (Si) and nickel (Ni) in the 2D and 3D simulations and post--process them with two different
radiation transport codes: {\it SNEC} \citep{2015ascl.soft05033M}, using an equillibrium--diffusion method and 
{\it SuperNu} \citep{2013ApJS..209...36W} using Implicit Monte Carlo (IMC) and Discrete Diffusion Monte Carlo (DDMC) methods under 
the assumption of local thermal equillibrium (LTE). We also present the first full 3D PISN synthetic spectra and LCs as a function of viewing angle
calculated by {\it SuperNu}.

This paper is organized as follows: Section~\ref{models} introduces the 1D, 2D and 3D simulations of the PISN model
used in this work ({\tt P250}), Section~\ref{analysis} presents synthetic LCs and spectra for all cases including the first
3D model spectra as a function of viewing angle in the literature and Section~\ref{disc} summarizes the implications 
of our results for the PISN mechanism.

\section{Multi--Dimensional PISN Models}\label{models}

Multidimensional (2D and 3D) simulations are necessary in order to assess the impact of hydrodynamic instabilities, such as RT, 
and mixing on the structure and radiative properties of PISNe. The first 2D PISN simulations using the {\it CASTRO} 
\citep{2010ApJ...715.1221A,2011ApJS..196...20Z,2013ApJS..204....7Z} code were presented by \citet{2011ApJ...728..129J} who
report little mixing between the O and He layers prior to shock breakout. \citet{2014ApJ...792...28C,2014ApJ...792...44C} also
explored the development of hydrodynamic instabilities in both PPISN and PISN in 2D and found that the upper and lower boundaries
of the O shell are unstable due to oxygen and helium burning shortly after core bounce. They also explored the role of a reverse
shock following SN shock breakout in driving the growth of RT instabilities. Rapidly rotating PISN progenitors simulated in
``2.5D'' show similar features \citep{2013ApJ...776..129C}.

A comprehensive study of mixing in PISN ejecta was presented by G17, who were
the first to perform a 3D PISN simulation using the adaptive mesh refinement (AMR) hydrodynamics code {\it FLASH} 
\citep{2000ApJS..131..273F,2012ApJS..201...27D}. G17 considered
a 250~$M_{\odot}$ PISN progenitor model with metallicity 0.07~$Z_{\odot}$ (model {\tt P250}) 
and computed with the stellar evolution code {\it GENEC} \citep{2012A&A...537A.146E,2013MNRAS.433.1114Y}. 
Strong radiatively--driven winds lead to complete
hydrogen envelope stripping for model {\tt P250} so that its pre--PISN mass is $\sim$~127~$M_{\odot}$ with
only $\sim$~2~$M_{\odot}$ of He retained in the outer layers. The hydrodynamic evolution, explosion and 
nucleosynthetic burning using 19 isotopes (the {\tt Aprox19} network in {\it FLASH}; \citealt{2000ApJS..126..501T})
are then simulated in 1D spherical, 2D cylindrical and 3D cartesian grids. 

Model {\tt P250} produces an energetic explosion ($\simeq 8.2 \times 10^{52}$~erg) synthesizing $\simeq 34$~$M_{\odot}$
of radioactive $^{56}$Ni. The original 1D and 2D/3D mass--weighted angular--averaged profiles were then post--processed
by the radiation--hydrodynamics code {\it STELLA} \citep{1998ApJ...496..454B} yielding a superluminous, slow--evolving
bolometric LC (Figure 15 of G17). A comparison of the PISN ejecta composition profiles in the angular--averaged 
2D and 3D profiles shown in Figure~8 of G17 suggests stronger mixing in 3D as compared to 2D, especially in the
interface between the Si and the O layer. Less extensive mixing is also seen in the Ni/Si interface. This chemical rearrangement is 
driven by the growth of the RT instability between layers of different composition in the PISN ejecta and is stronger
in 3D as expected \citep{2014PlPhR..40..451K}.

As mentioned in G17, the RT instabilities were still growing at the end of the simulations. In order to reach the final mass fraction distributions, 
we needed to extend the evolution with {\it FLASH} until mixing has effectively ceased for this work. To accomplish this, we mapped the 1D, 2D, and 3D post--shock data 
($r < 2.5 \times 10^{10}$~cm) from G17 onto larger grids to facilitate expansion of the shock to $r \sim 1 \times 10^{11}$~cm. 
The pre--shock regions of the grid were filled with the densities, temperatures, and mass fractions (effectively uniform) from the {\it GENEC} progenitor model. 
As was done in G17 to extend simulations for input into {\it STELLA}, we modify the densities and temperatures to eliminate a jump discontinuity that is 
caused by the edge density decreasing in the original explosion simulations. 
However, we choose the radius for this to be $7.56 \times 10^{10}$~cm as the
fixed radius used in G17 lies outside of our domain of $1 \times 10^{11}$~cm.

For the 1D simulation we use the same refinement criteria and bounding resolutions ($1.3 \times 10^{8}$~cm and $6.5 \times 10^{7}$~cm) from G17. 
For 2D and 3D we use a nested refinement structure with distinct refinement regions that decrease in refinement outwards. 
We turn off AMR so that the nested grid structure is static throughout the simulations, as in G17. Due to computational limitations, 
we could not continue the 3D simulation with the same maximum resolution as in G17 ($3.25 \times 10^{7}$~cm). We do, however, map the
data exactly from the final {\it FLASH} checkpoint files of G17 onto the corresponding interior spatial regions of the larger grids of the extended {\it FLASH} 
simulations. Then, we perform derefinement in the inner refinement regions (a cylinder in 2D and a cube in 3D) so that they merge with their 
neighboring refinement regions. Thus, evolution of the extended {\it FLASH} simulations begin with a maximum refinement of $6.5 \times 10^{7}$~cm in their 
interior refinement regions. Care was taken to ensure that the compositional interfaces never cross the boundaries of the inner refinement regions 
so that all mixing occurs at maximum resolution. Focusing on the compositional interfaces then, they form in the original simulations from 
G17 before mixing during their expansion. Then, they suddenly experience a decrease in refinement level and continue to mix and expand 
during the extended simulation. To understand the effect of the derefinement between simulations, we also computed the 2D simulation without such derefinement. 
The results of the mixing in the 2D simulation remain practically unchanged, therefore, the mixing in our extended simulations is converged at the resolution used. 
We halted the simulations when the SN shock wave was very near the edge of the grid ($1 \times 10^{11}$~cm). 
These simulations were performed using {\it XSEDE} computing resources \citep{xsede} together with {\it Los Alamos National Lab} ({\it LANL}) 
institutional computing resources.

Then, using the data from the final states of the {\it FLASH} {\tt P250} simulation in 1D, 2D and 3D geometry described in the previous step,
we prepared seven profiles for input into the equillibrium--radiation diffusion 
code {\it SNEC} \citep{2015ascl.soft05033M} and the IMC--DDMC radiation transport code {\it SuperNu} \citep{2013ApJS..209...36W}.
These profiles include the direct {\it FLASH} output from the 1D {\tt P250} simulation (model {\tt 1D}), and three mass--weighted angular averages each for the 
2D and 3D {\tt P250} simulation outputs. One such profile is created from the average over all angles (models {\tt 2D\_AA} and {\tt 3D\_AA}), 
and two profiles from restricting the average over one degree and (one degree)$^{2}$ in the 2D and 3D cases, respectively. 
The profiles from the restricted averages are meant to represent two opposing viewing angles in which the Si is most effectively 
mixed outward (models {\tt 2D\_MO} and {\tt 3D\_MO}) or inward (models {\tt 2D\_MI} and {\tt 3D\_MI}) at the Si/O interface. 
For these restricted averages, all cells whose centroids lay within the range(s) of angles are included. 

The averaging proceeded by separating the included cells into radial bins spanning from zero to $1 \times 10^{11}$~cm 
(corresponding to the location of the SN shock wave). 
For the full mass averages, these bins were uniformly spaced in the 2D cases at $1.5 \times 10^{8}$, and in the 3D case at $5 \times 10^{8}$. 
The reason for this was to ensure that enough cells were included in the average for each bin and in 3D there were fewer 
cells to work with in the stellar envelope (where the simulation resolution was lower than in the 2D simulation). Both bin sizes are more than 
sufficient for capturing the radial widths of features caused by mixing and were also used for the restricted averages in the
region where mixing occurred. However, since even fewer points are available in a restricted average, we needed slightly larger 
bin sizes for the inner core and envelope regions, both of which have
an effectively uniform composition.

After construction of the radial grids the state variables were averaged for each bin (using the individual cell masses for weighting). 
All of the state variables tracked in the {\it FLASH} simulation, save for internal energy and pressure (which can be derived from the 
density and temperature via the Timmes EoS; \citealt{2000ApJS..126..501T}), were averaged. 
Finally, the profiles resulting from the restricted averages were smoothed using a running--line smoother with a span of three 
\citep{Hast:Tibs:1990}. This was done to eliminate noise due to a relative paucity of points in comparison to the full averages.

\setcounter{table}{0}
\begin{deluxetable*}{lccccccccccccc}
\tabletypesize{\footnotesize}
\tablecaption{Basic properties of the PISN ({\tt P250}) models presented in this work.}
\tablehead{
\colhead {Model} &
\colhead{$E_{\rm SN}$~(B)} &
\colhead {$M_{\rm Ni}$~($M_{\odot}$)} &
\colhead{$\theta^{\dagger}$~($^\circ$)} &
\colhead {$\phi^{\dagger}$~($^\circ$)} &
\colhead {$L_{\rm max,D}$~($10^{44}$~erg~s$^{-1}$)} &
\colhead {$t_{\rm max,D}$} &
\colhead {$L_{\rm max}$~($10^{44}$~erg~s$^{-1}$)} &
\colhead {$t_{\rm max}$} &
\colhead {$t_{\rm Si/O}$} &
\colhead {$t_{\rm Ni/Si}$} &
\\}
\startdata
\hline
{\tt 1D}            &  81.9 & 34.0 & --          & --          & 1.06 & 185.7 & 1.79 & 159.5 & 181.0 & 216.4 \\
{\tt 2D\_AA}    &  81.9 & 34.0 & --          & --          & 1.06 & 190.0 & 1.81 & 178.7 & 185.0 & 220.4 \\
{\tt 2D\_MI}     &  81.9 & 34.0 & 18--19  & --          & 1.09 & 188.7 & 1.84 & 176.0 & 200.7 & 224.3 \\
{\tt 2D\_MO}   &  81.9 & 34.0 & 37--38  & --          & 1.05 & 182.8 & 1.78 & 168.4 & 169.2 & 216.4 \\
{\tt 3D\_AA}    &  81.8 & 33.8 & --          & --          & 1.05 & 185.4 & 1.74 & 165.3 & 181.0 & 220.4 \\
{\tt 3D\_MI}     &  81.8 & 33.8 & 17--18  & 73--74  & 0.94 & 186.9 & 1.57 & 171.7 & 246.0 & 251.9 \\
{\tt 3D\_MO}   &  81.8 & 33.8 & 21--22  & 69--70   & 1.00 & 171.2 & 1.61 & 165.9 & 153.5 & 202.7 \\
{\tt 3D\_FULL}$^{\ddagger}$ &  81.8 & 33.8 & --          & --          & -- & -- & 1.23 & 185.0 & N/A & N/A  \\
\enddata 
\tablecomments{$^{\dagger}$~$\theta$ and $\phi$ correspond to the polar and the azimuthal angle, accordingly, in 3D spherical coordinates. 
For the 2D models, the polar angle $\theta$ is the only coordinate used. The peak luminosity ($L_{\rm peak,D}$) and time of peak luminosity 
($t_{\rm peak,D}$) values as computed in {\it SNEC} using equilibrium--diffusion radiation transport are also quoted.
$^{\ddagger}$ The values quoted for $L_{\rm max}$ and $t_{\rm max}$ in model {\tt 3D\_FULL} correspond to a viewing angle of $\Omega \simeq$~0$^\circ$
(``edge--on'' view). These values vary only by a small amount for different choices of $\Omega$. We do not provide $t_{\rm Si/O}$ and $t_{\rm Ni/Si}$ estimates 
for the {\tt 3D\_FULL} model because there is not a unique time when the photosphere crosses the Si/O and Ni/Si compositional interfaces
in the 3D simulations due to the large extent of the RT mixing. All timescales are expressed in units of days.
\label{T1}}
\end{deluxetable*}



Table~\ref{T1} details the basic properties of the 1D profiles adopted from the 1D, 2D and 3D model {\tt P250} simulations
for all cases (full angular--average, Si ``mixed inwards'' and Si ``mixed outwards''). Our models include a full 3D profile of model {\tt P250}
({\tt 3D\_FULL}) that was directly mapped in the 3D homologous grid of {\it SuperNu} yielding LCs and spectra as a function of
viewing angle (Section~\ref{supernu3D}). Table~\ref{T1} also lists the peak luminosity ($L_{\rm max, D}$ and $L_{\rm max}$ for synthetic
LCs computed in {\it SNEC} and {\it SuperNu} respectively), time to peak luminosity ($t_{\rm max, D}$ and $t_{\rm max}$ for {\it SNEC}
and {\it SuperNu} respectively) and the timescales corresponding to the phase when the SN photosphere crosses the Si/O interface
($t_{\rm Si/O}$) and the Ni/Si interface ($t_{\rm Ni/Si}$) of the ejecta, all measured in units of days.

\begin{figure*}
\gridline{\fig{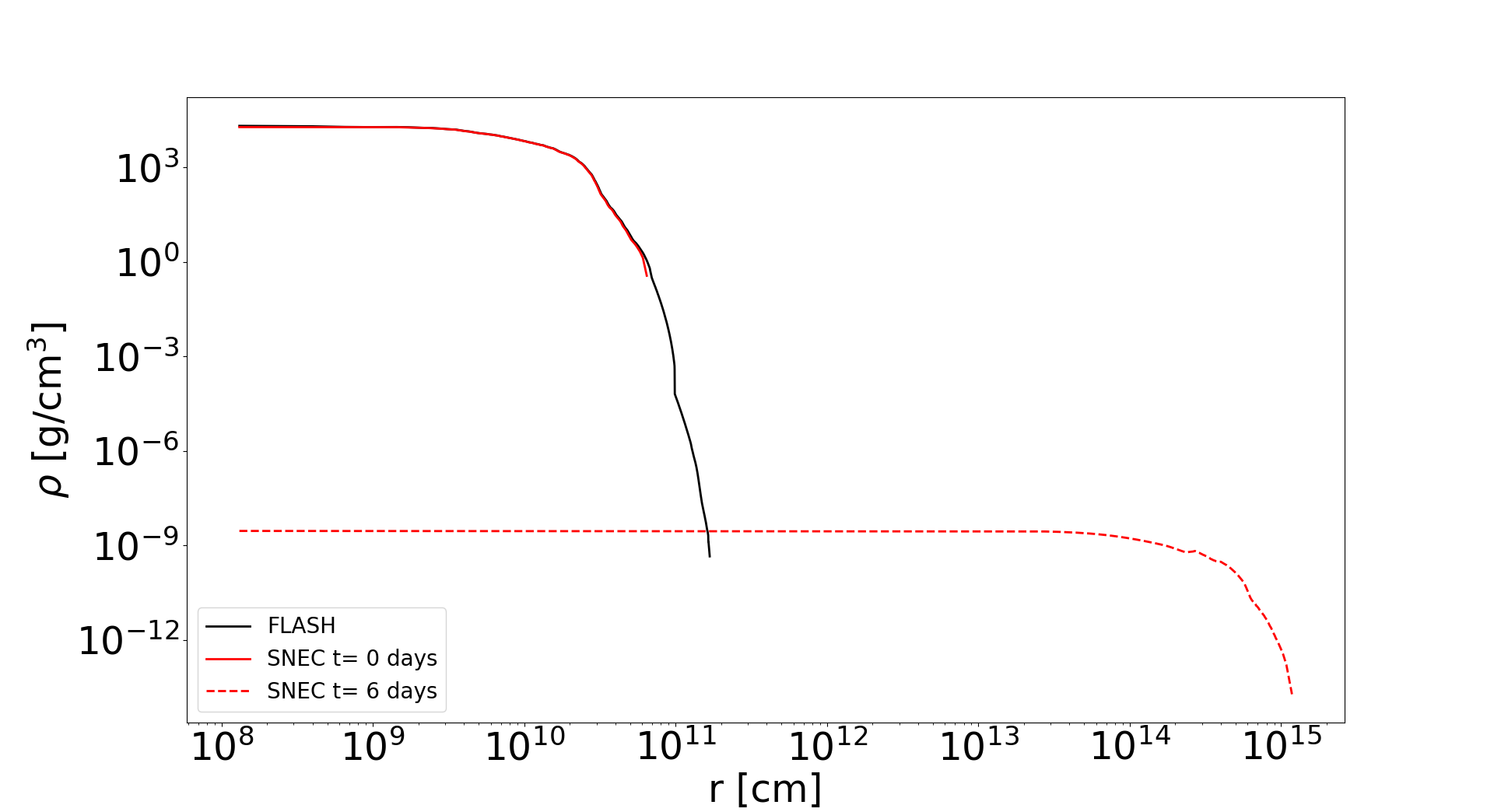}{0.5\textwidth}{}
          \fig{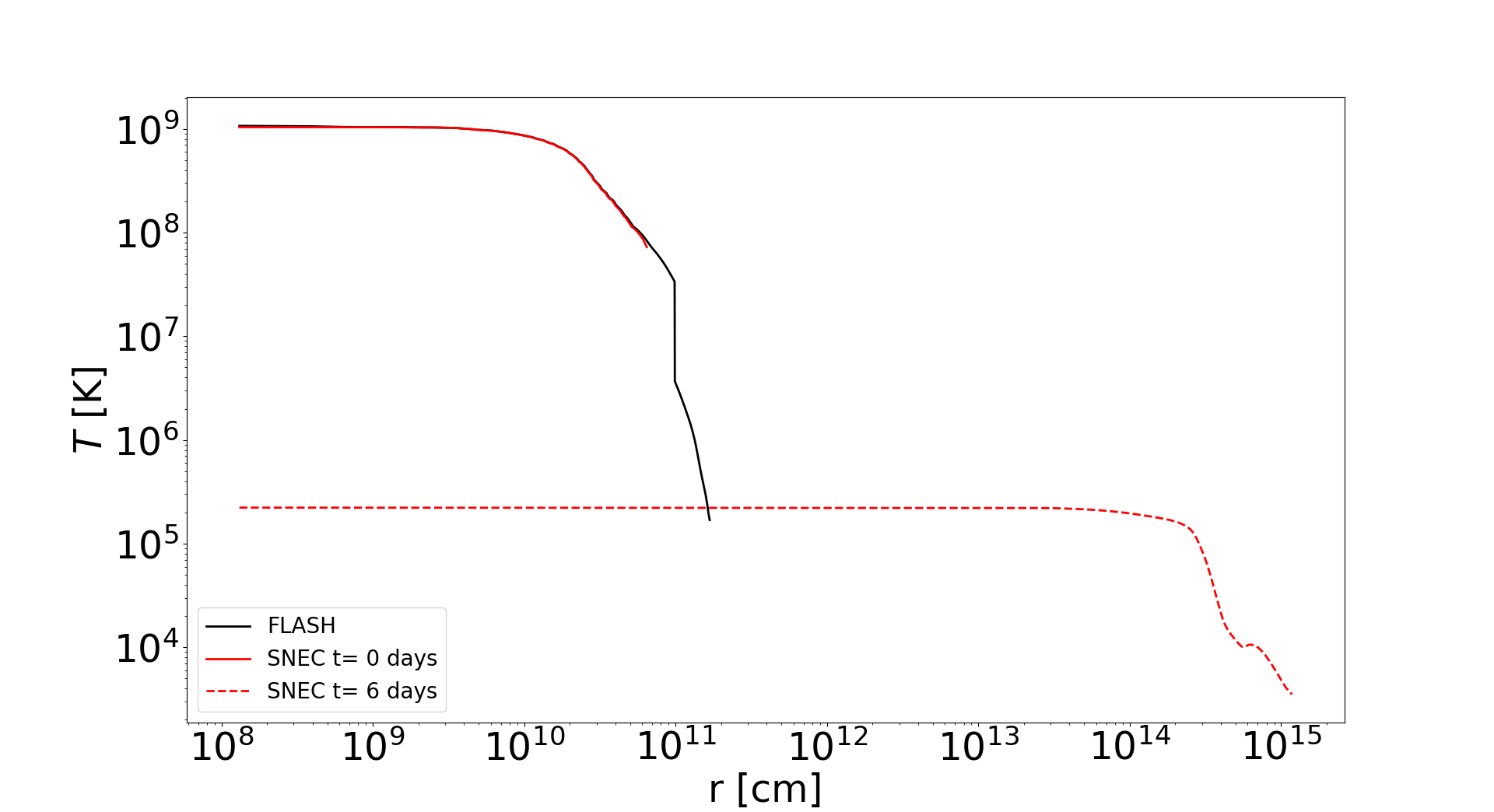}{0.5\textwidth}{}
          }
\gridline{\fig{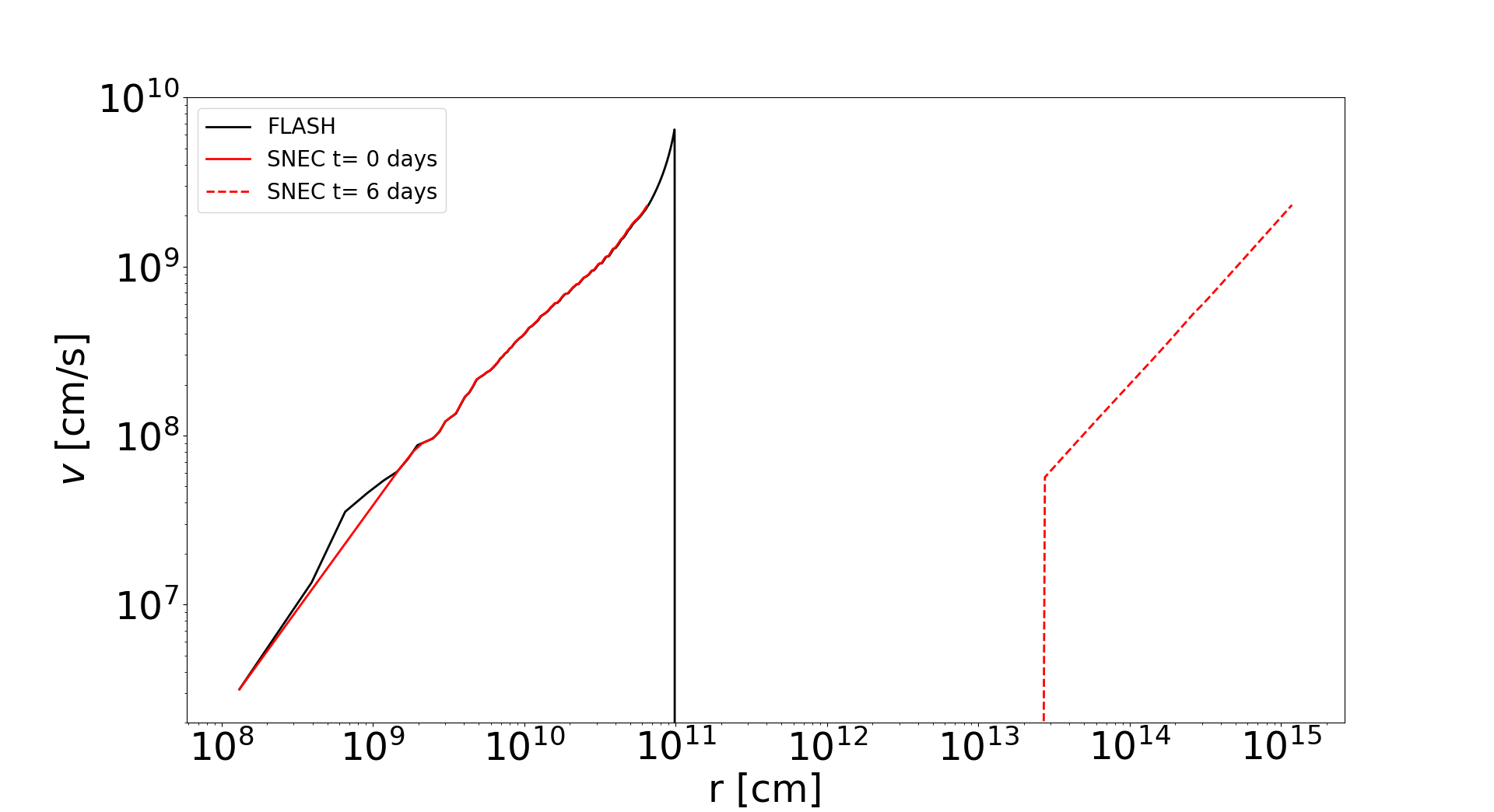}{0.5\textwidth}{}
         \fig{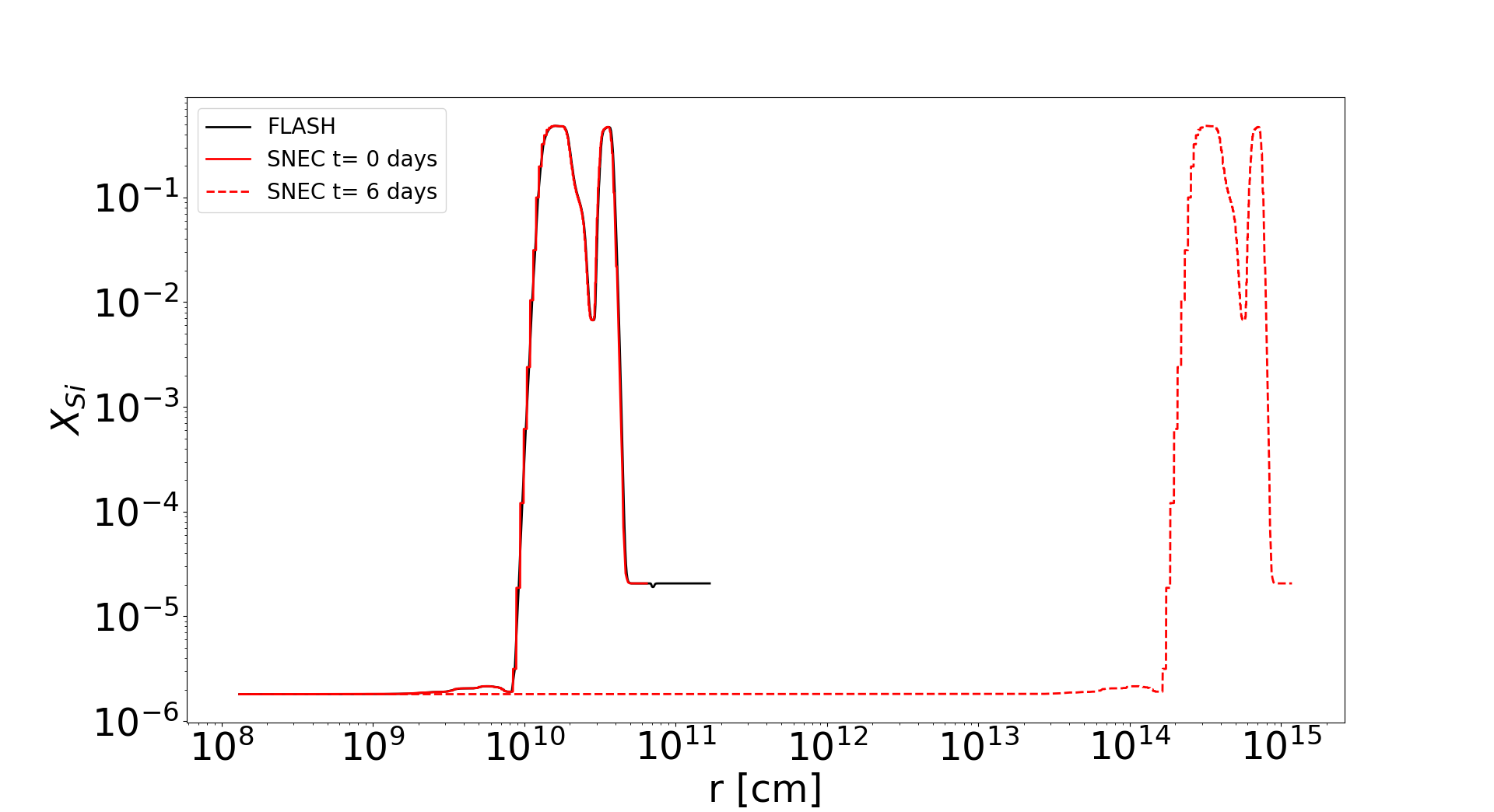}{0.5\textwidth}{}
         }
\caption{Density ($\rho$; {\it upper left panel}), temperature ($T$; {\it upper right panel}), velocity ($v$; {\it lower left panel}) and Si mass fraction ($v$; {\it lower right panel}) profiles
of the {\tt 1D} PISN model. Black curves correspond to the original {\it FLASH} profiles and red curves to the same profiles after mapping to the grid of {\it SNEC}. The dashed red curve
shows the {\it SNEC} profiles after 6 days of homologous hydrodynamic evolution.
\label{Fig:flash_snec}}
\end{figure*}

\begin{figure*}
\gridline{\fig{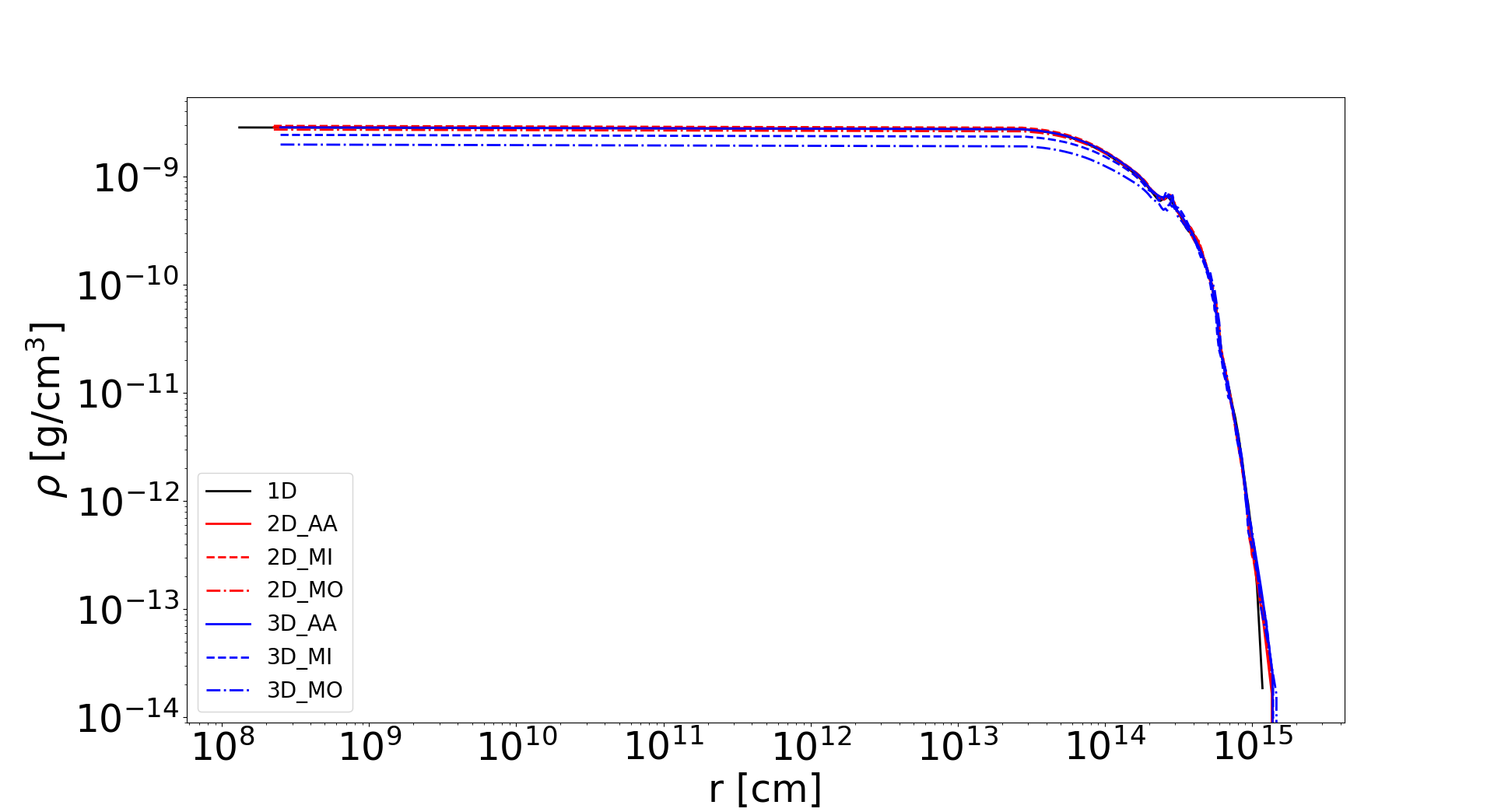}{0.5\textwidth}{}
          \fig{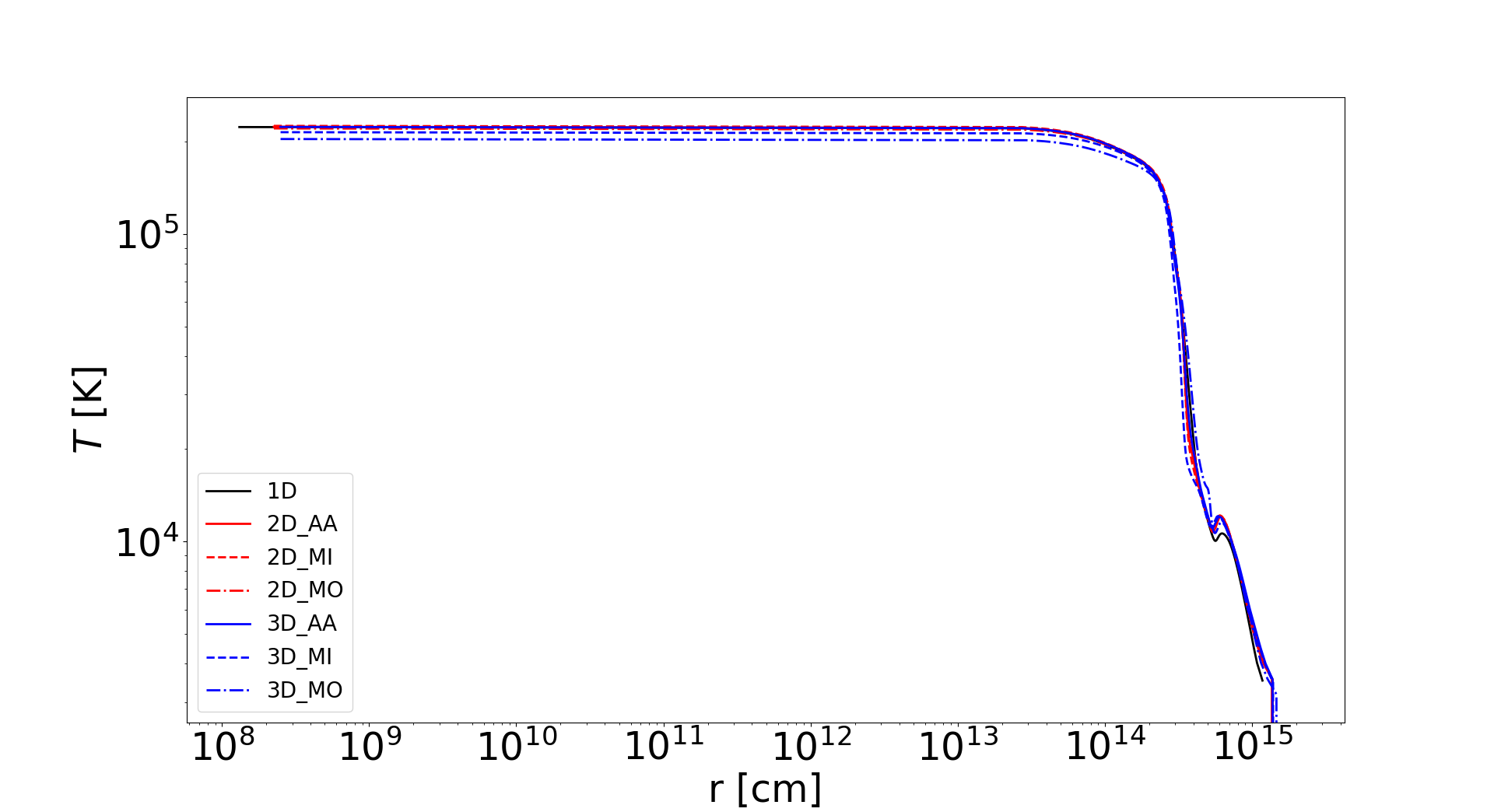}{0.5\textwidth}{}
          }
\caption{Comparisons of density ($\rho$; {\it left panel}) and temperature ($T$; {\it right panel}) profiles for models: {\tt 1D} (black curves), {\tt 2D\_AA, 2D\_MI, 2D\_MO} (red curves)
and  {\tt 2D\_AA, 2D\_MI, 2D\_MO} (blue curves) at $t =$~6~days in the {\it SNEC} grid.
\label{Fig:snec6d}}
\end{figure*}

\begin{figure*}
\gridline{\fig{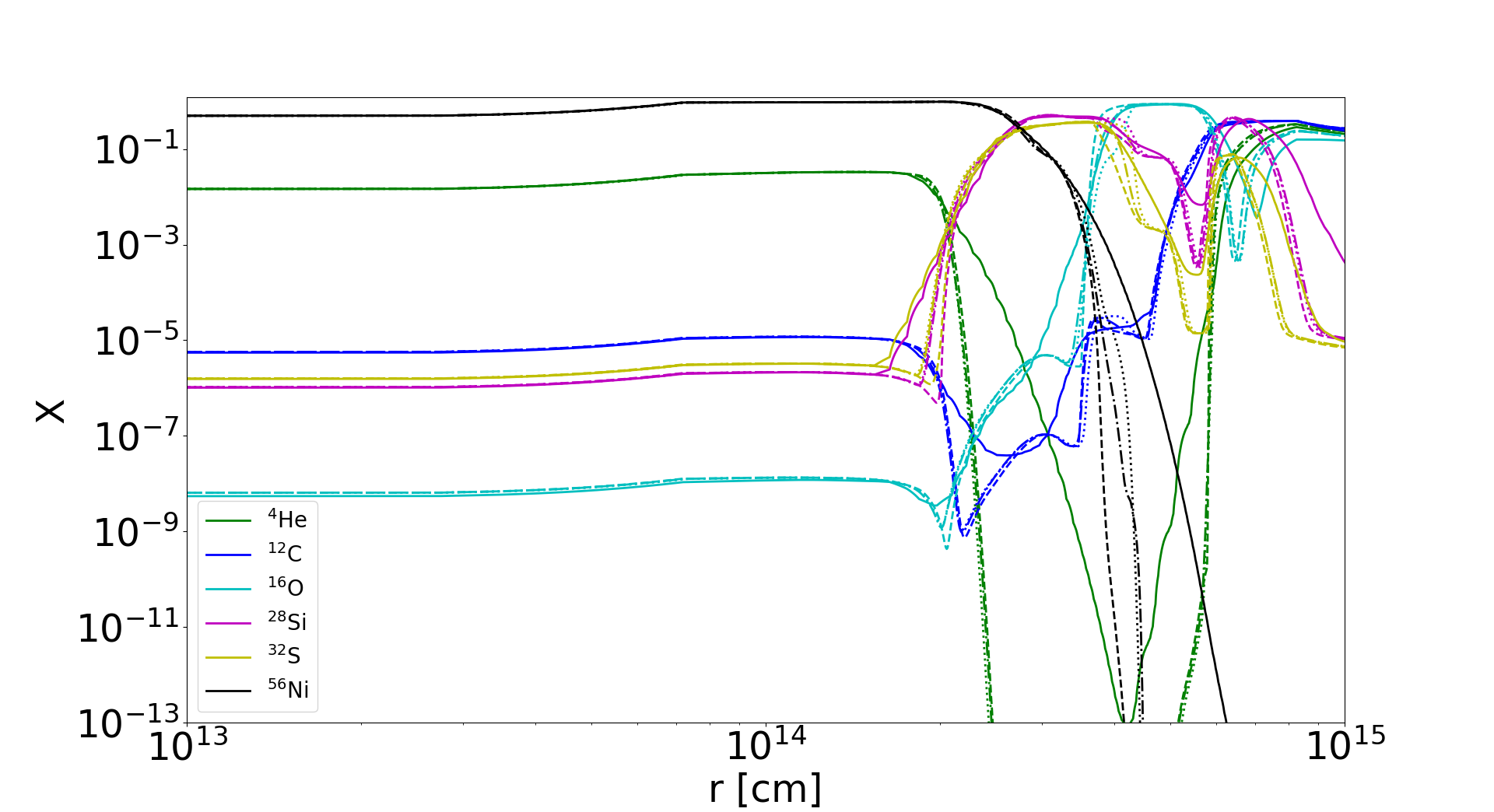}{0.5\textwidth}{}
          \fig{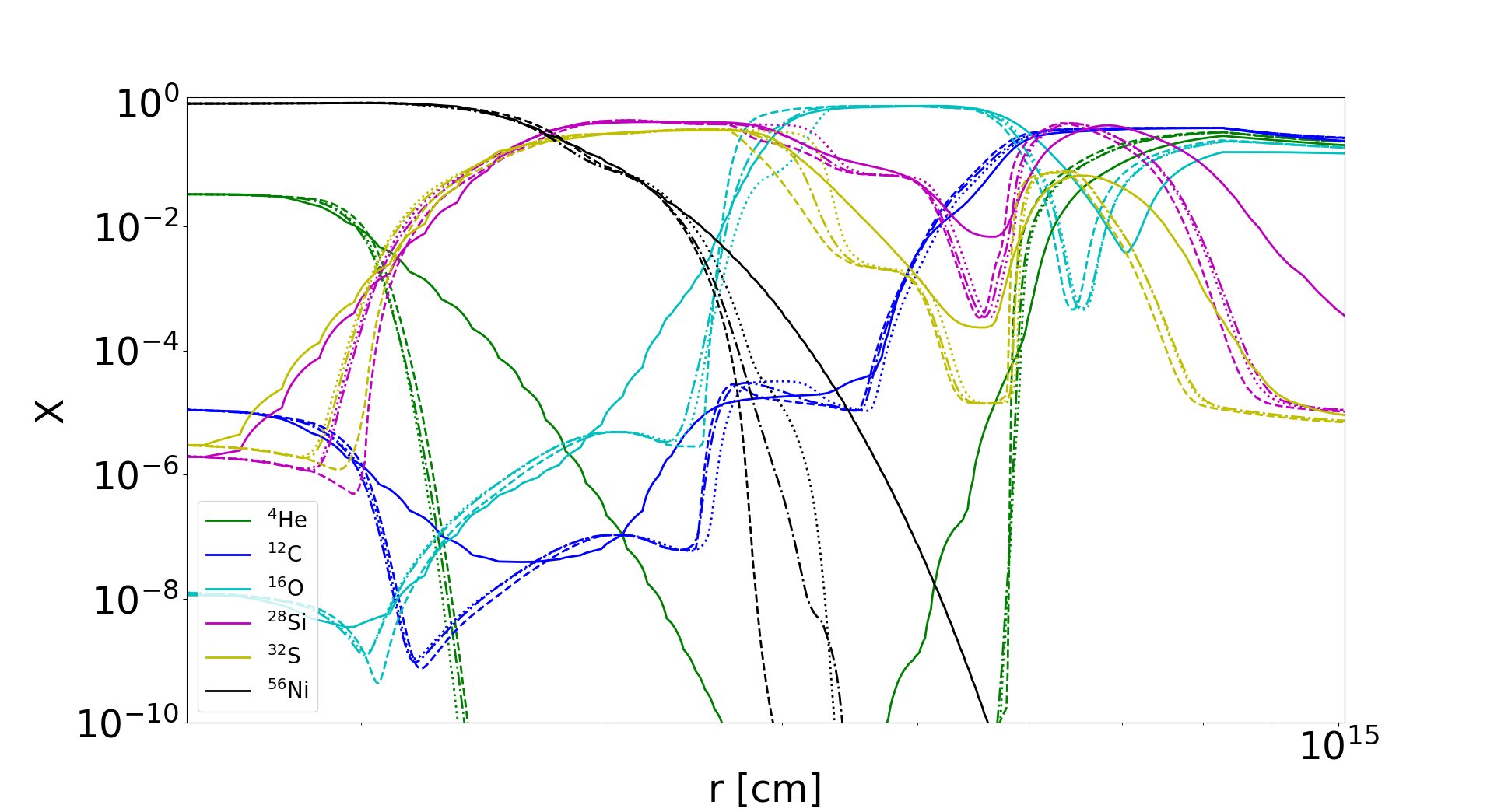}{0.5\textwidth}{}
          }
\gridline{\fig{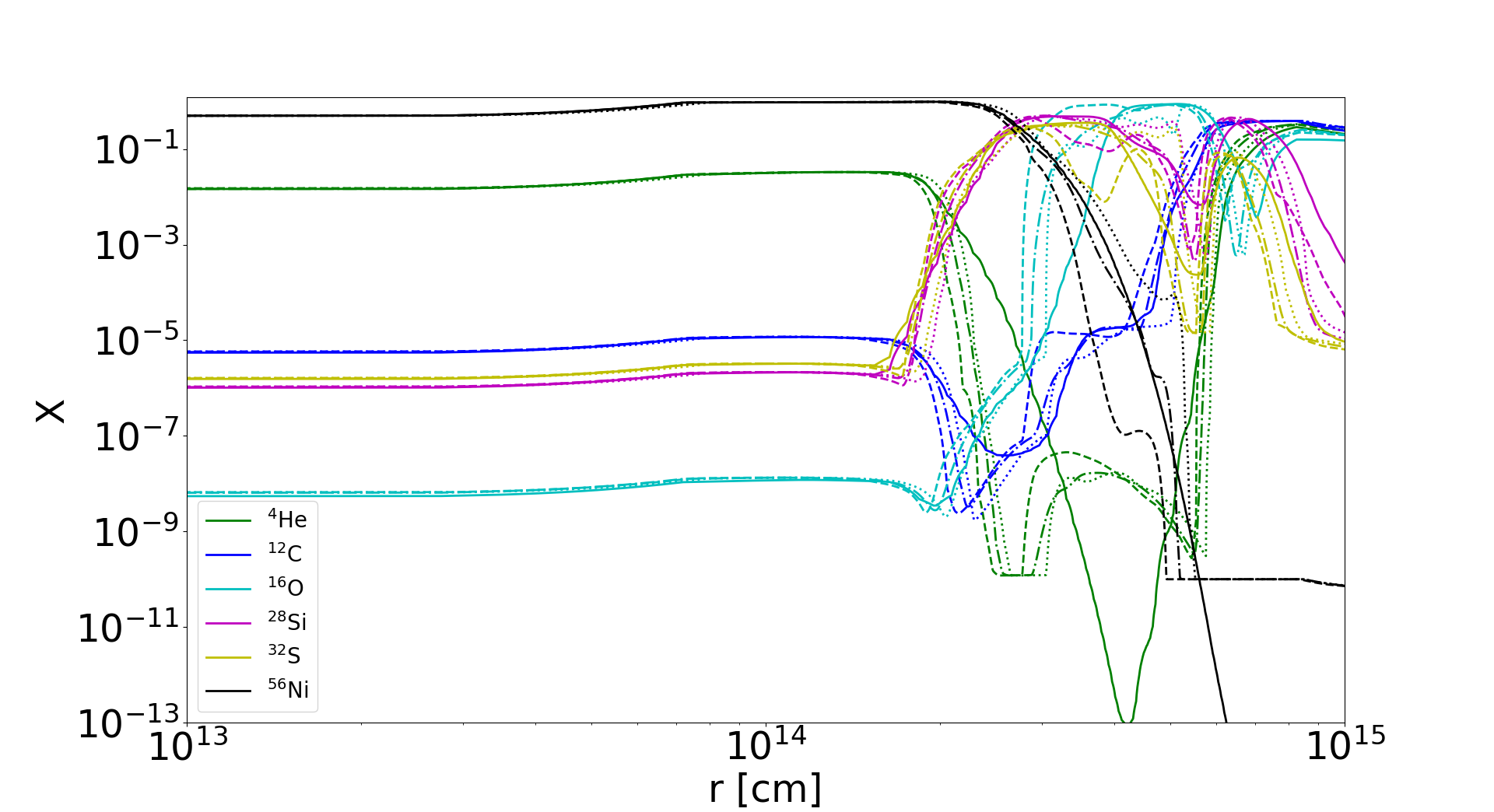}{0.5\textwidth}{}
         \fig{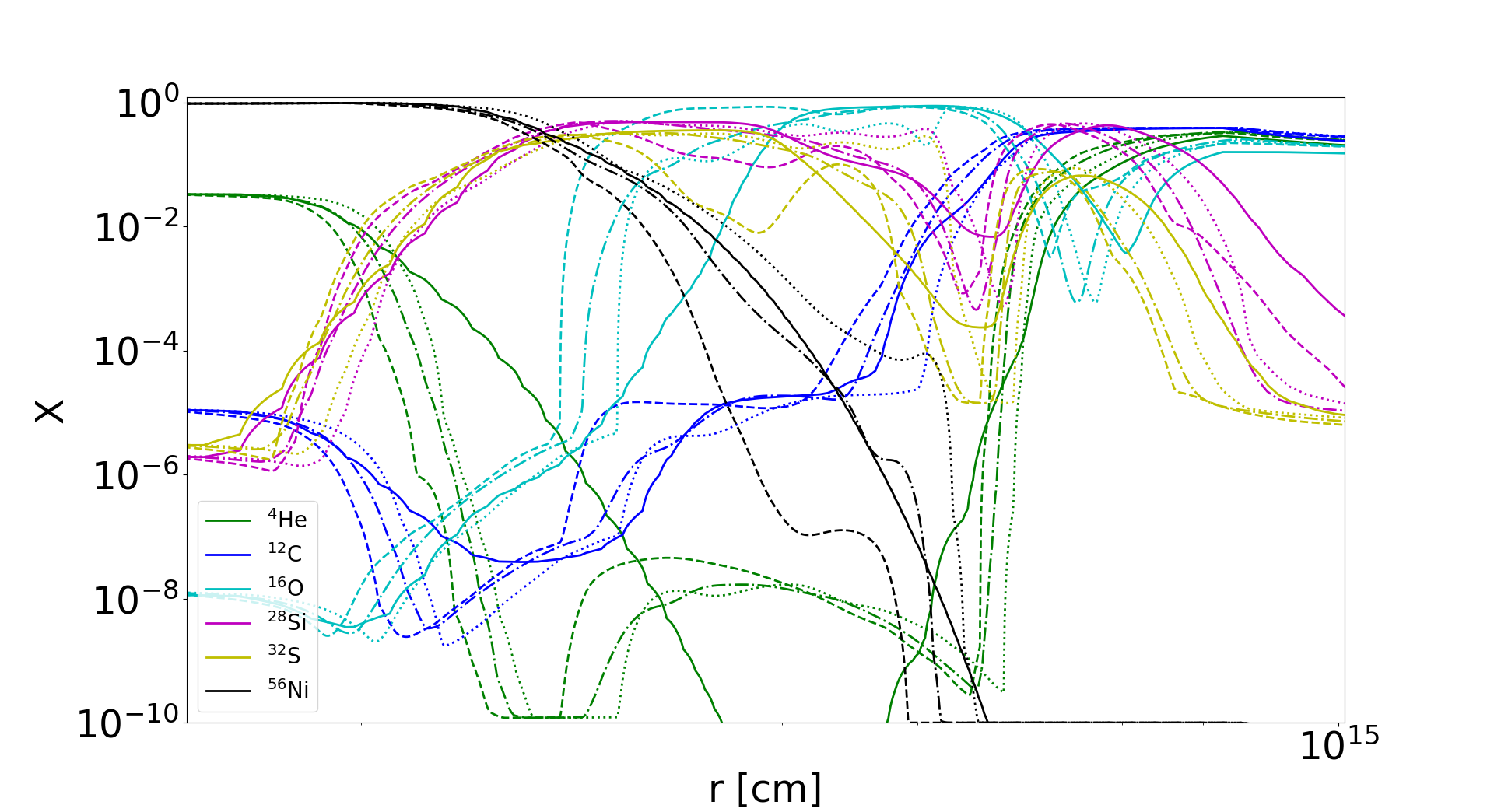}{0.5\textwidth}{}
         }
\caption{Comparison of the abundance profiles at $t =$~6~days in the {\it SNEC} grid. The upper panels correspond to the {\tt 1D} and the {\tt 2D\_AA, 2D\_MI, 2D\_MO} models and
the lower panels to the {\tt 1D} and the {\tt 3D\_AA, 3D\_MI, 3D\_MO} models. In each case, the right panels show a zoom--in to the regions of high Si/O and Ni/Si mixing.
The solid curves represent the {\tt 1D}, the dot--dashed curves the angular--averaged, the dashed curves the Si ``mixed--inwards'' and the dotted curves the Si
``mixed--outwards'' models.
\label{Fig:snec6dcomp}}
\end{figure*}

\begin{figure}
\begin{center}
\includegraphics[angle=0,width=9cm]{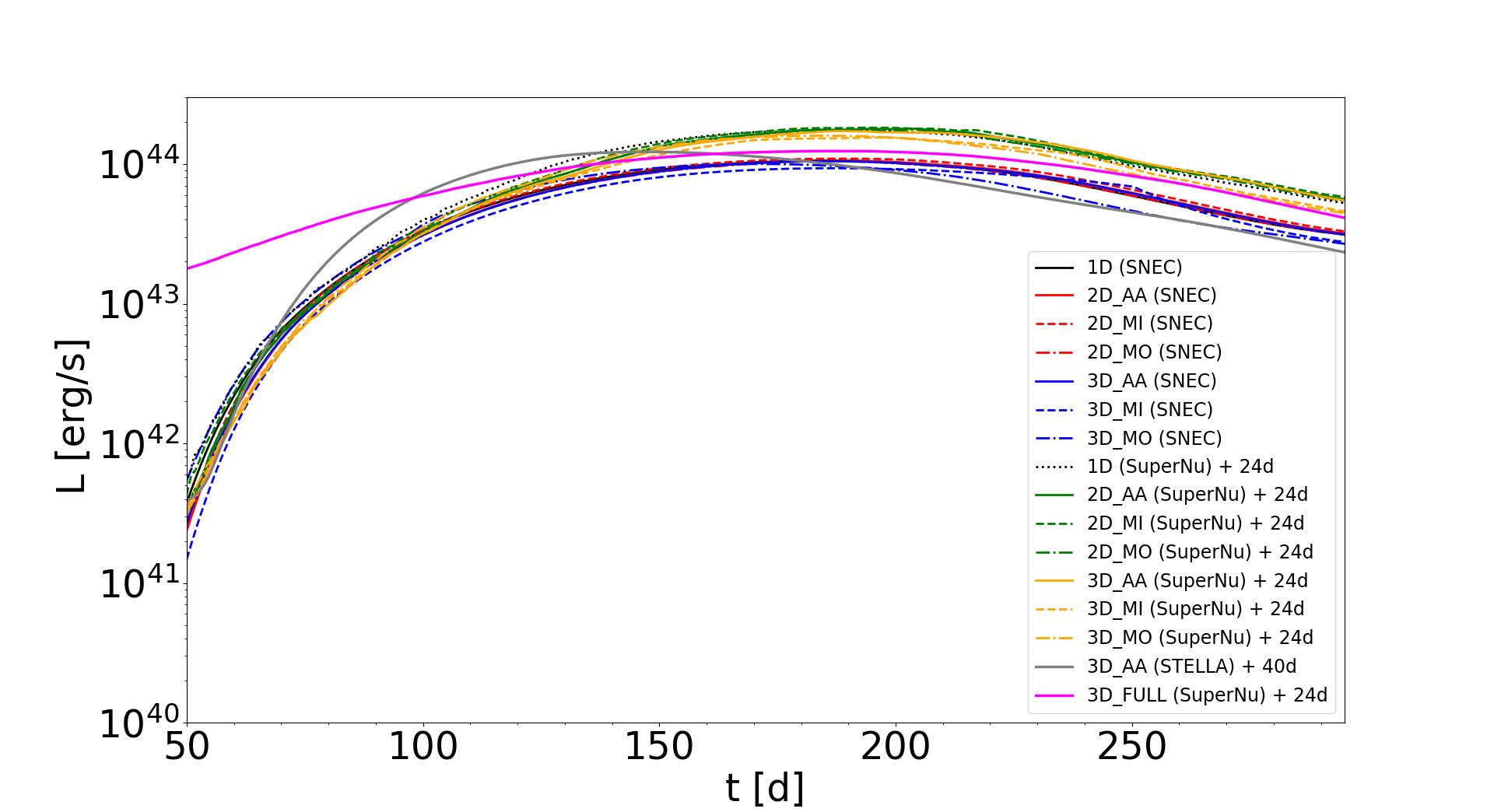}
\caption{Synthetic bolometric lightcurves from {\it SNEC} and {\it SuperNu} for the PISN profiles used in this work. The bolometric LC for the angular--averaged 3D {\tt P250} model
computed with {\it STELLA} and presented in \citep{2017ApJ...846..100G} is also shown for comparison.
\label{Fig:lcs_log}}
\end{center}
\end{figure}

\section{PISN SYNTHETIC LIGHTCURVES AND SPECTRA}\label{analysis}

\subsection{{\it Hydrodynamic Expansion and Radiation Diffusion with SNEC}}\label{snec}

The seven profiles adopted from the 1D, 2D and 3D {\tt P250} model simulations were all at a phase
corresponding to SN shock radius of $1 \times 10^{11}$~cm and therefore still interior to the stellar envelope of the progenitor star
(with radius $\simeq 1.68 \times 10^{11}$~cm). In order to compute synthetic spectra with {\it SuperNu}, the main objective of this work, 
we need to provide an input profile that corresponds to a phase shortly after SN shock breakout when the ejecta start to expand homologously. 
For this reason we decided to follow the post--shock breakout evolution of each of these profiles out to $\sim$~6~days using 
the hydrodynamics solver included in the {\it SNEC} code. {\it SNEC} is a spherically-symmetric Lagrangian 
radiation-hydrodynamics code designed to follow supernova explosions through the envelope of their progenitor star, 
produce bolometric (and approximate multi--color) LC predictions, and provide input to spectral synthesis codes for spectral modeling.
As such, {\it SNEC} also enables us to perform comparisons with {\tt P250} LCs computed with {\it SuperNu} and {\it STELLA}.

Figure~\ref{Fig:flash_snec} shows the density, temperature, velocity and Si mass fraction of the {\tt 1D} profile at the time of mapping
into the Langragian grid of {\it SNEC} ($t =$~0~d) and at $t =$~6~d when the SN shock reaches a radius of $\sim 10^{15}$~cm. It can be
seen that upon mapping to {\it SNEC} a monotonically--increasing homologous velocity profile is recovered and the abundance profile
of Si (and all other species) remains frozen since nuclear burning has ceased long ago, shorty after the PISN explosion. A comparison
of the density and temperature profiles at $t =$~6~d for all 7 profiles of model {\tt P250} discussed here is shown in Figure~\ref{Fig:snec6d}.
The abundance profiles for six main species present in the SN ejecta ($^{4}$He, $^{12}$C12, $^{16}$O, $^{28}$Si, $^{32}$S, $^{56}$Ni) at $t =$~6~d  
are shown in Figure~\ref{Fig:snec6dcomp} including zoomed--in views in regions of high mixing around the Ni/Si and Si/O interfaces. The effect
of higher mixing in the {\tt 3D\_MI} and {\tt 3D\_MO} models as compared to their 2D counterparts is clearly seen.

At the beginning of the {\it SNEC} simulation, the SN profiles are already in a phase of homologous expansion with high velocities
($\gtrapprox$~50,000~km~s$^{-1}$) so we did not have to impose an artificial thermal bomb or a piston explosion to advance the SN evolution.
The {\it SNEC} calculations include heating by the radioactive decay of $^{56}$Ni that is the dominant power--input in LCs of PISNe
and realistic material opacities adopted from the {\it OPAL} \citep{1996ApJ...464..943I} database. The
{\tt P250} model produces a $^{56}$Ni yield of $~\sim$~34~$M_{\odot}$ enabling a superluminous bolometric LC 
($L_{\rm peak,D} \simeq 1.05 \times 10^{44}$~erg~s$^{-1}$, where ``D'' stands for ``diffusion'' implying the radiation diffusion calculation
as done in {\it SNEC}). 

Figure~\ref{Fig:lcs_log} presents the synethic LCs for all models used in this work as computed with {\it SNEC} (solid black, red
and blue curves), {\it SuperNu} (dotted black, green and orange curves) as well as a comparison with the {\tt 3D\_AA} {\tt P50} model LC 
calculated with {\it STELLA} (Figure 15 of G17). The peak luminosity agrees within a factor of $\sim$~2 between the three codes but
the timescale to rise to peak luminosity ($t_{\rm max}$) is considerably shorter for the {\it STELLA} calculation compared to that found by
{\it SNEC} and {\it SuperNu} (discussed in the next paragraph). 
This could be due to a variety of reasons including grid resolution, the implementation used to simulate radiation transfer
and, more importantly, line opacity values. Investigating these differences is beyond the scope of this paper but we refer to \citet{2017MNRAS.464.2854K}
for a thorough discussion on radiation transfer code--to--code comparison specifically applied to PISN models where the same discrepancy 
is found between {\it STELLA} and the Monte Carlo radiation transport code {\it SEDONA} \citep{2006ApJ...651..366K}.
The model LCs exhibit minor differences between the angular--averaged, the Si ``mixed--inwards'' and the Si ``mixed--outwards'' cases. The most
notable difference is seen for the {\tt 3D\_MI} and {\tt 3D\_MO} models that both reach lower peak luminosities (especially during the post--maximum phase)
as compared to all other models. This behavior is consistent between the {\it SNEC} and {\it SuperNu} results and illustrates 
the effect of mixing on the total radiated flux from the PISN explosion. This is discussed in more detail in the following paragraph.

\subsection{{\it 1D Synthetic Spectra with SuperNu}}\label{supernu1D}

{\it SuperNu} is utilized to compute time series of synthetic spectra, synthetic LCs and assess how 
the radiative properties of PISNe are affected by mixing captured in multidimensional simulations (G17, and this paper).
The $t =$~6~day homologous {\it SNEC} profiles for all cases were mapped into the Langragian grid of the IMC--DDMC radiation trasfer code {\it SuperNu} 
\citep{2013ApJS..209...36W} in 1D spherical geometry.
{\it SuperNu} simulates time--dependent radiation transport in LTE with matter. It applies the methods of Implicit Monte Carlo (IMC) \citep{Fleck} 
and Discrete Diffusion Monte Carlo (DDMC) \citep{Densmore} for static or homologously expanding spatial grids.  
The radiation field affects material temperature but does not affect the motion of the fluid. Multi--group absorption opacity data from hydrogen up to cobalt are included in
addition to line data for bound--bound opacities from \citep{1995KurCD..23.....K}. {\it SuperNu} features an improved implementation of opacity regrouping to non--contiguous 
frequency groups that leads to enhanced performance and computational efficiency \citep{Wollaeger2014}. This is further enhanced by the capacity of {\it SuperNu} to be run on many compute cores in
parallel mode. In addition, {\it SuperNu} has the capacity to solve the radiative transport and diffusion equations in 1D spherical, 2D cylindrical and 3D spherical, cylindrical or cartesian 
geometries yielding viewing angle--dependent synthetic LCs and spectra.

\begin{figure*}
\gridline{\fig{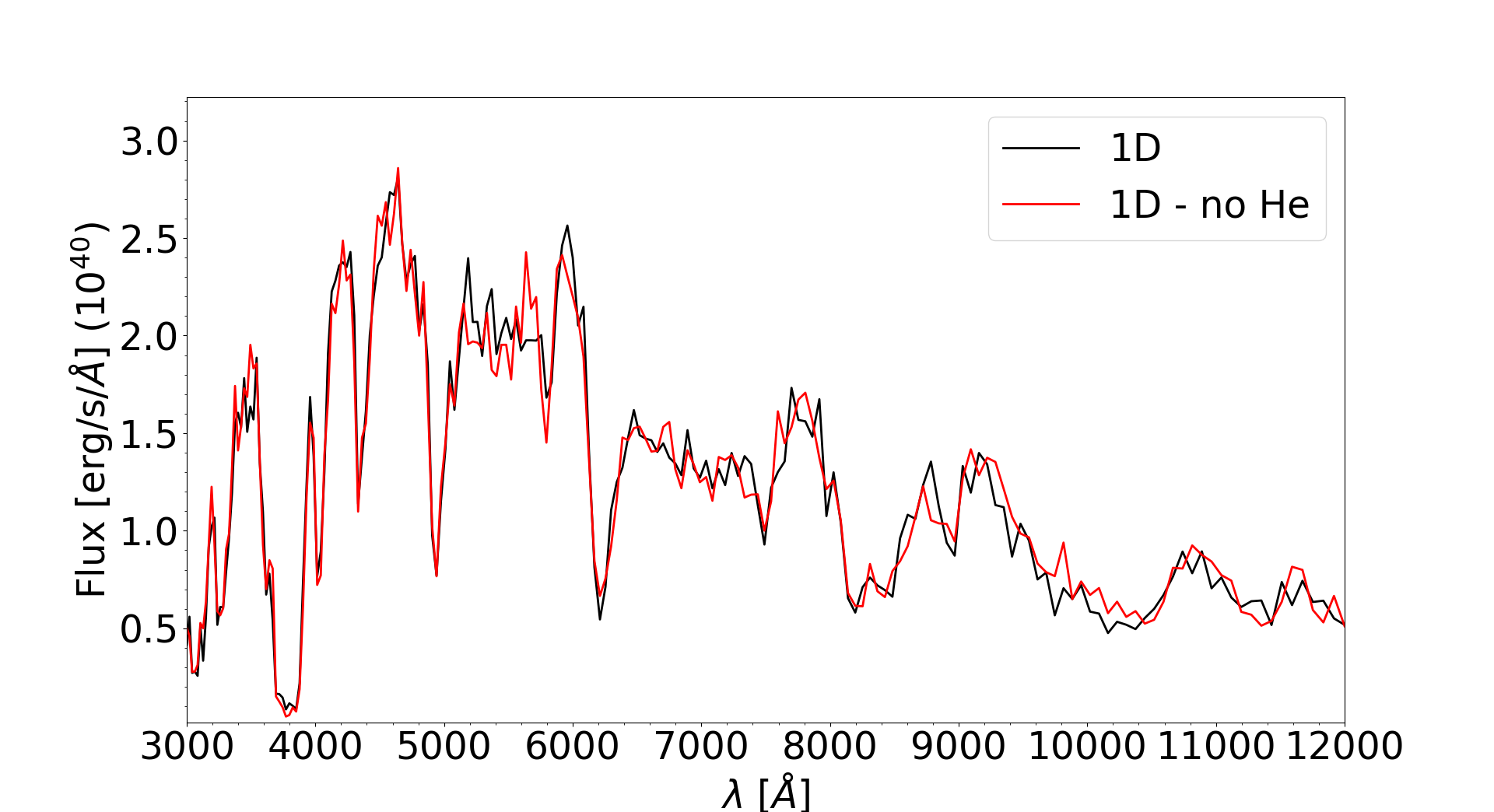}{0.5\textwidth}{}
          \fig{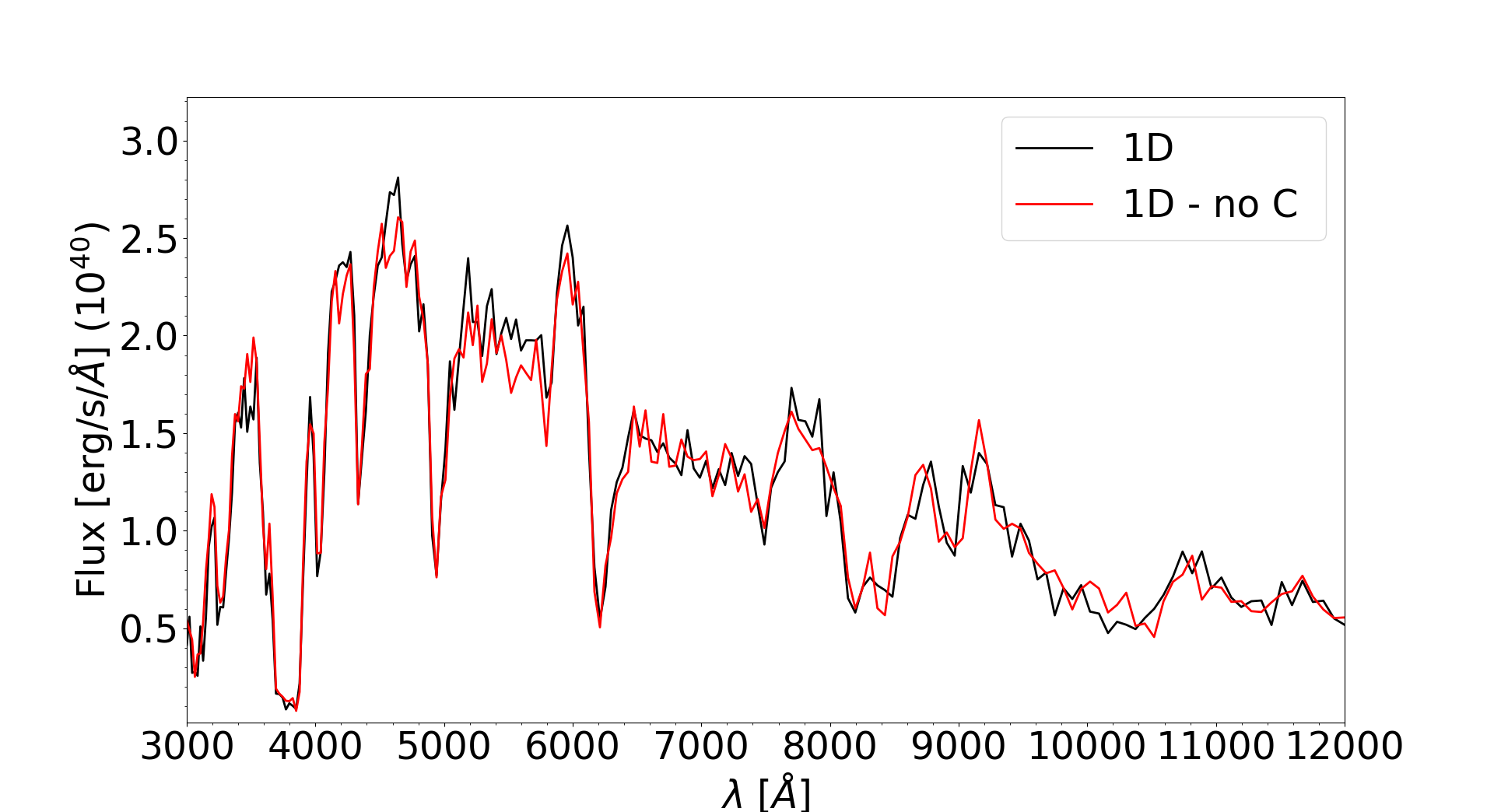}{0.5\textwidth}{}
          }
\gridline{\fig{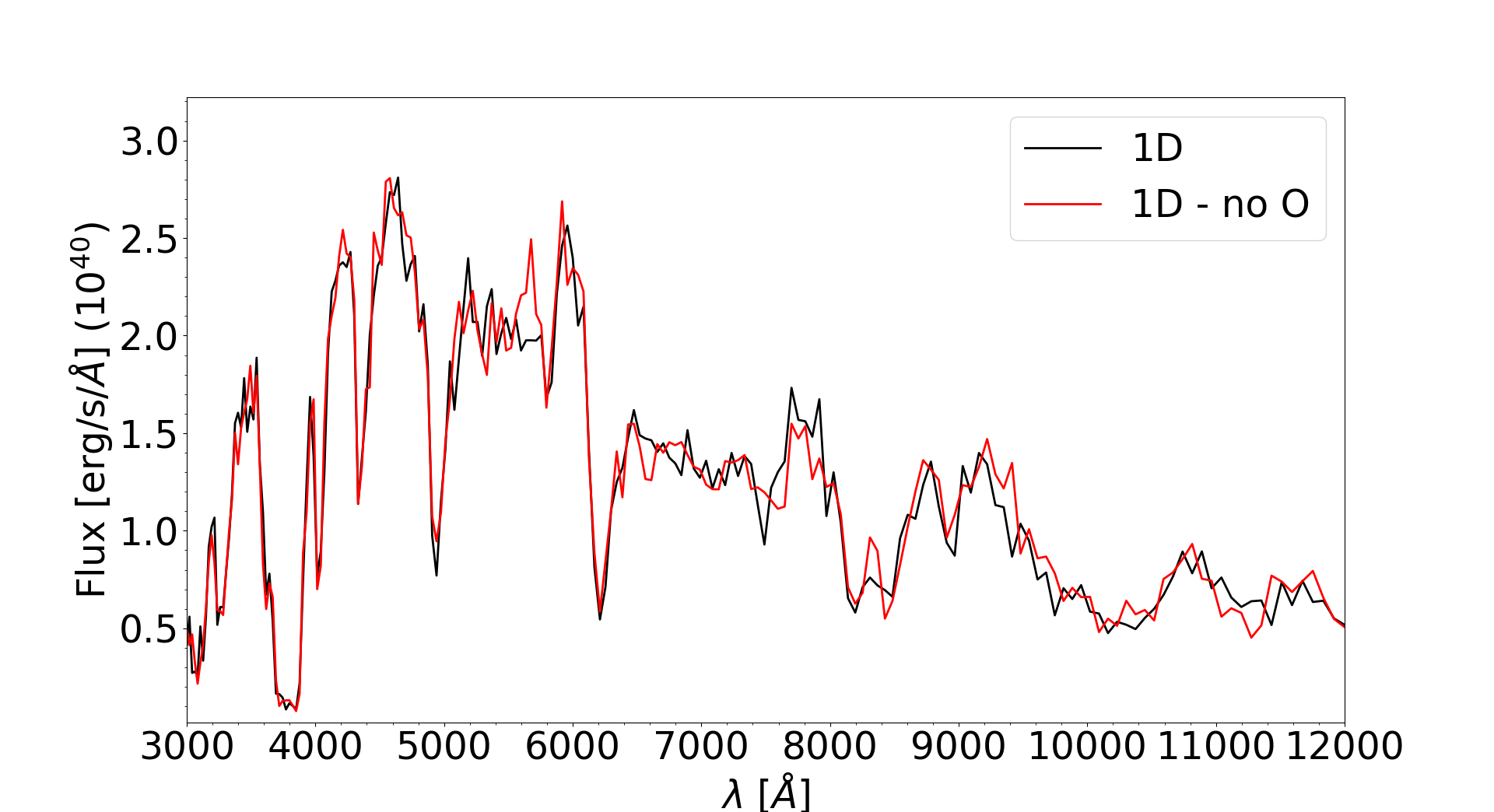}{0.5\textwidth}{}
         \fig{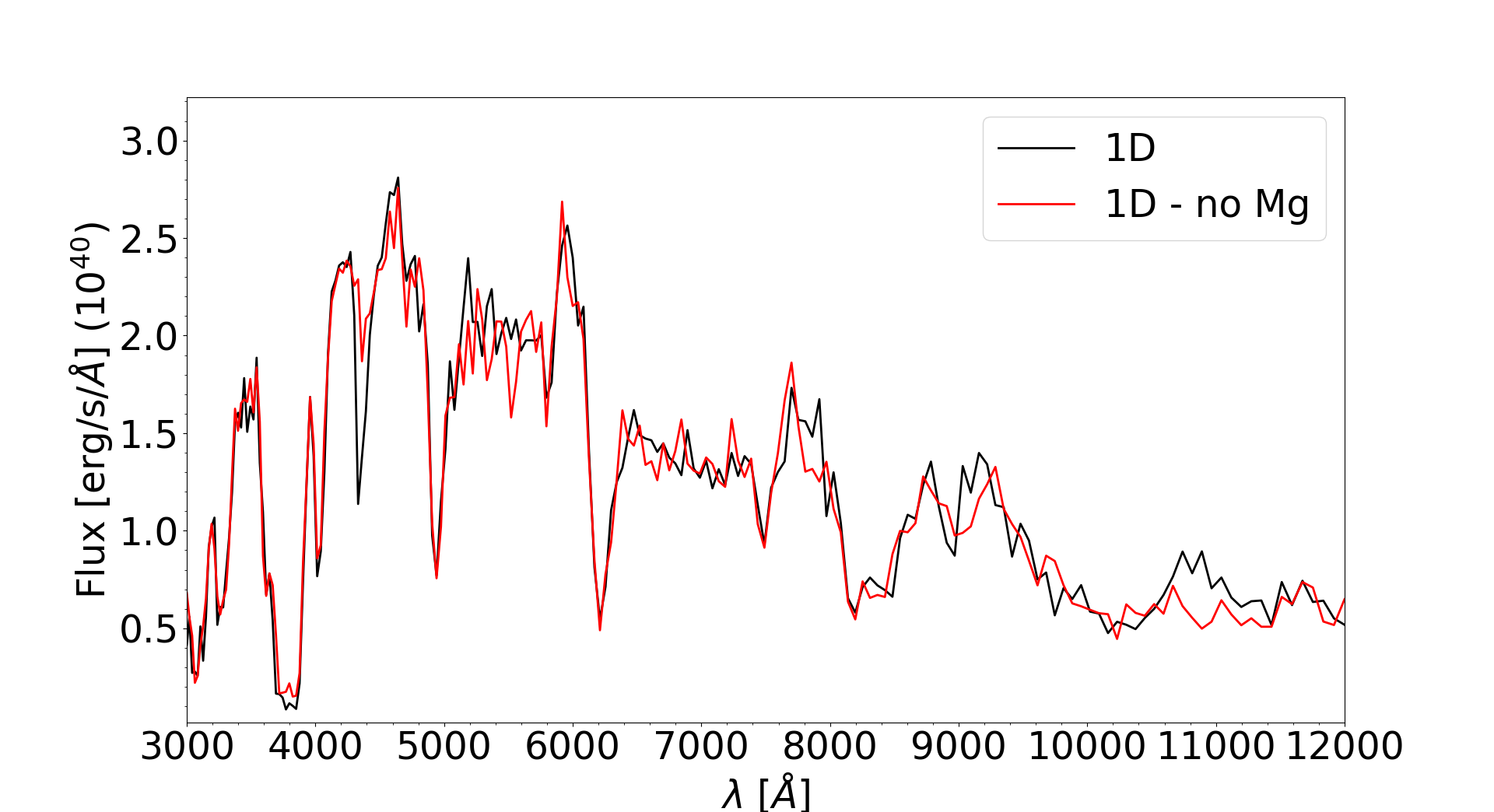}{0.5\textwidth}{}
         }
\gridline{\fig{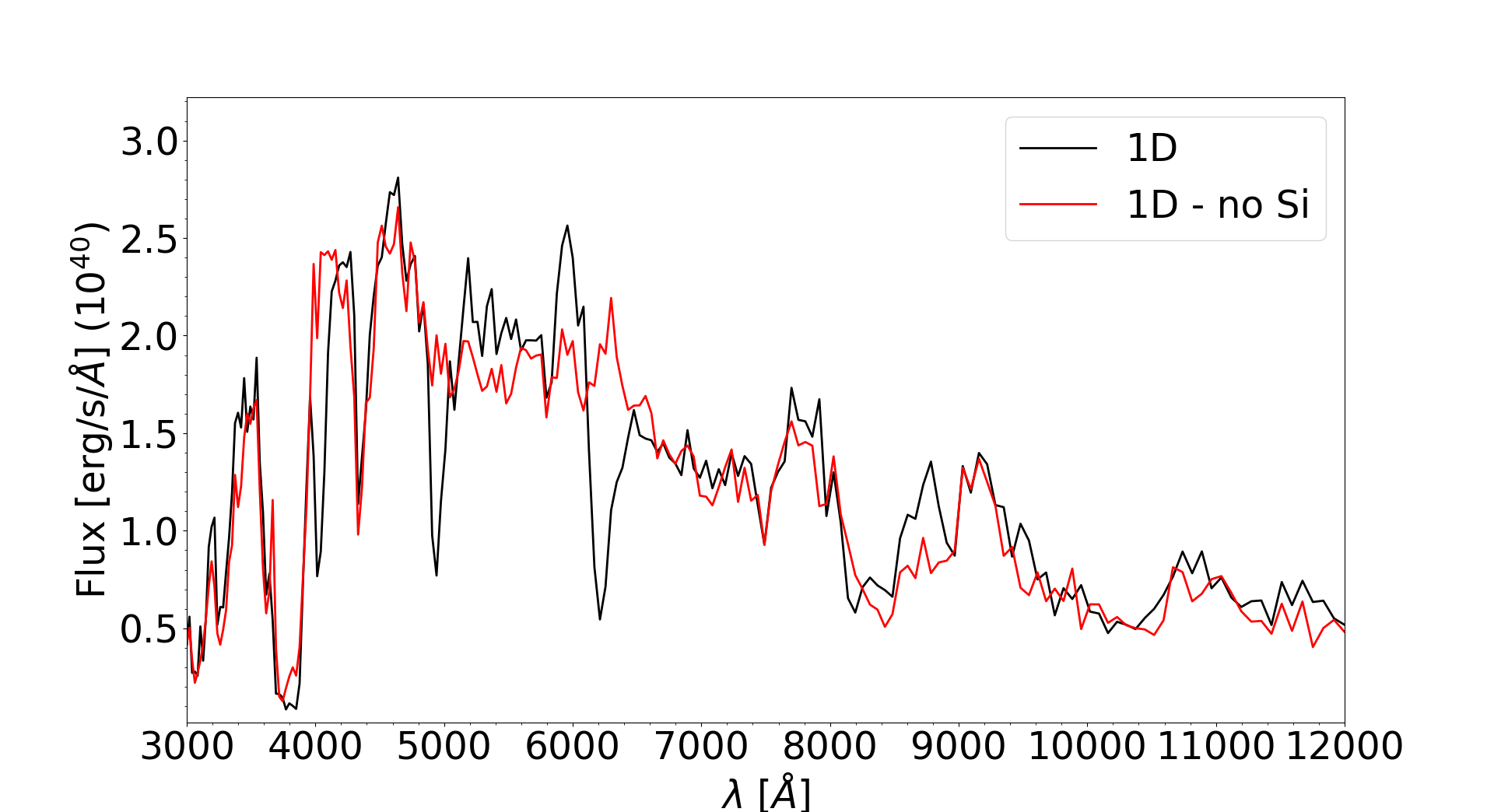}{0.5\textwidth}{}
         \fig{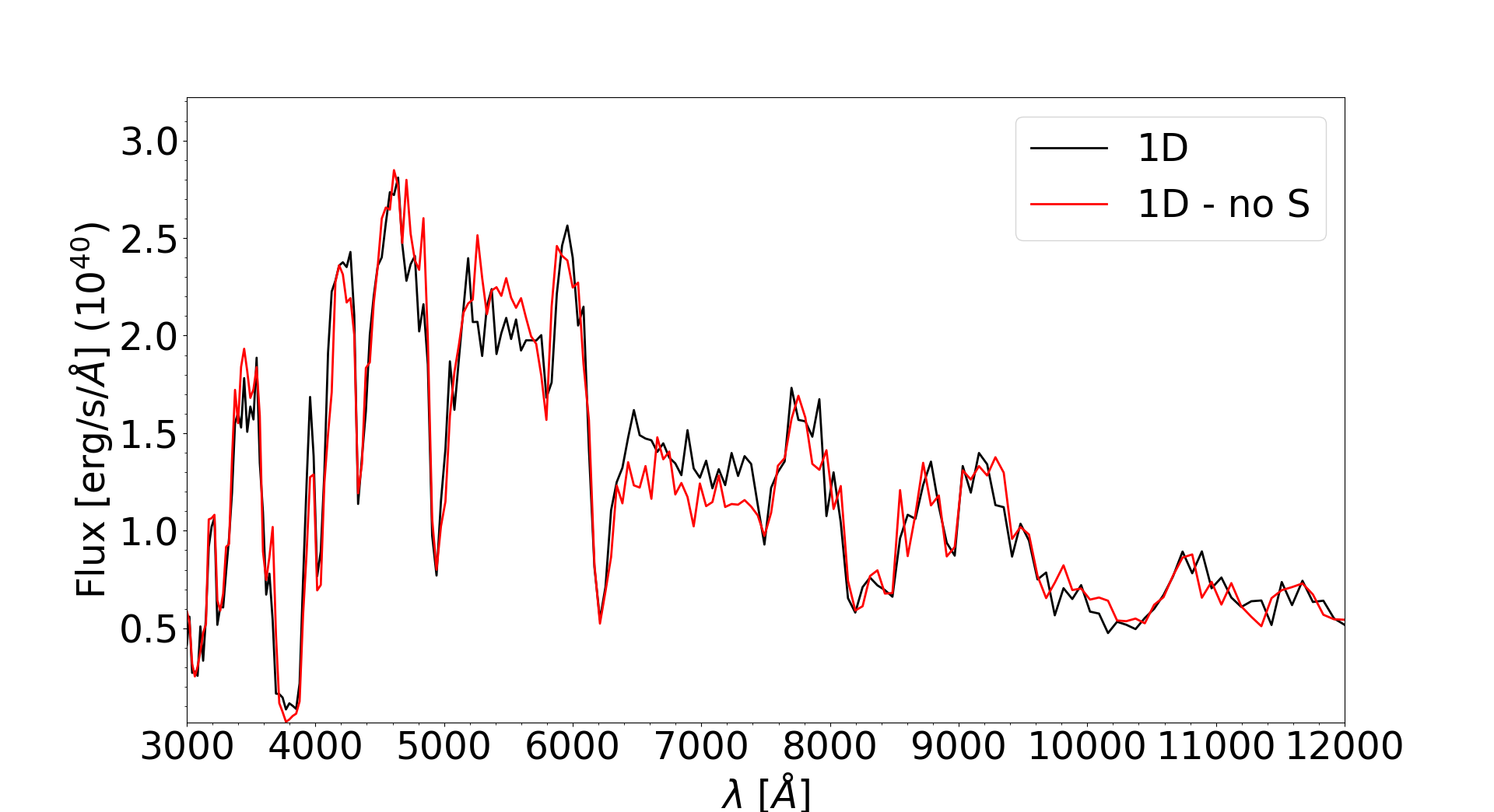}{0.5\textwidth}{}
          }
\caption{Synthetic {\it SuperNu} spectrum at peak luminosity for the {\tt 1D} model after subtracting contributions from isotopes of He ({\it upper left panel}), C ({\it upper right panel}), 
O ({\it middle left panel}), Mg ({\it middle right panel}), Si ({\it lower left panel}) and S ({\it lower right panel}). In each panel, the black curve corresponds 
to the full spectrum and the red curves to the spectrum after subtracting line transition data from all isotopes of the corresponding atom.  
\label{Fig:lineid}}
\end{figure*}

The synthetic spectrum of the {\tt 1D} model at peak luminosity is shown in Figure~\ref{Fig:lineid} including comparisons with runs that omitted isotopes for certain atomic species 
(He, C, O, Mg, Si or S) in order to identify dominant spectral features. It can be seen that the main features are due to intermediate mass elements (IMEs; most 
prominently Si followed by O, Mg and S), and that the spectrum is very similar to that of regular Type Ia SN events \citep{2000asqu.book.....C}. We confirmed this by using the
{\it SuperNova IDentification} ({\it SNID}) code \citep{2007ApJ...666.1024B} to compare the peak {\tt 1D} model spectrum to thousands of spectral templates of observed SNe
of different types yielding templates of Type Ia SNe as the best matches.


\begin{figure*}
\gridline{\fig{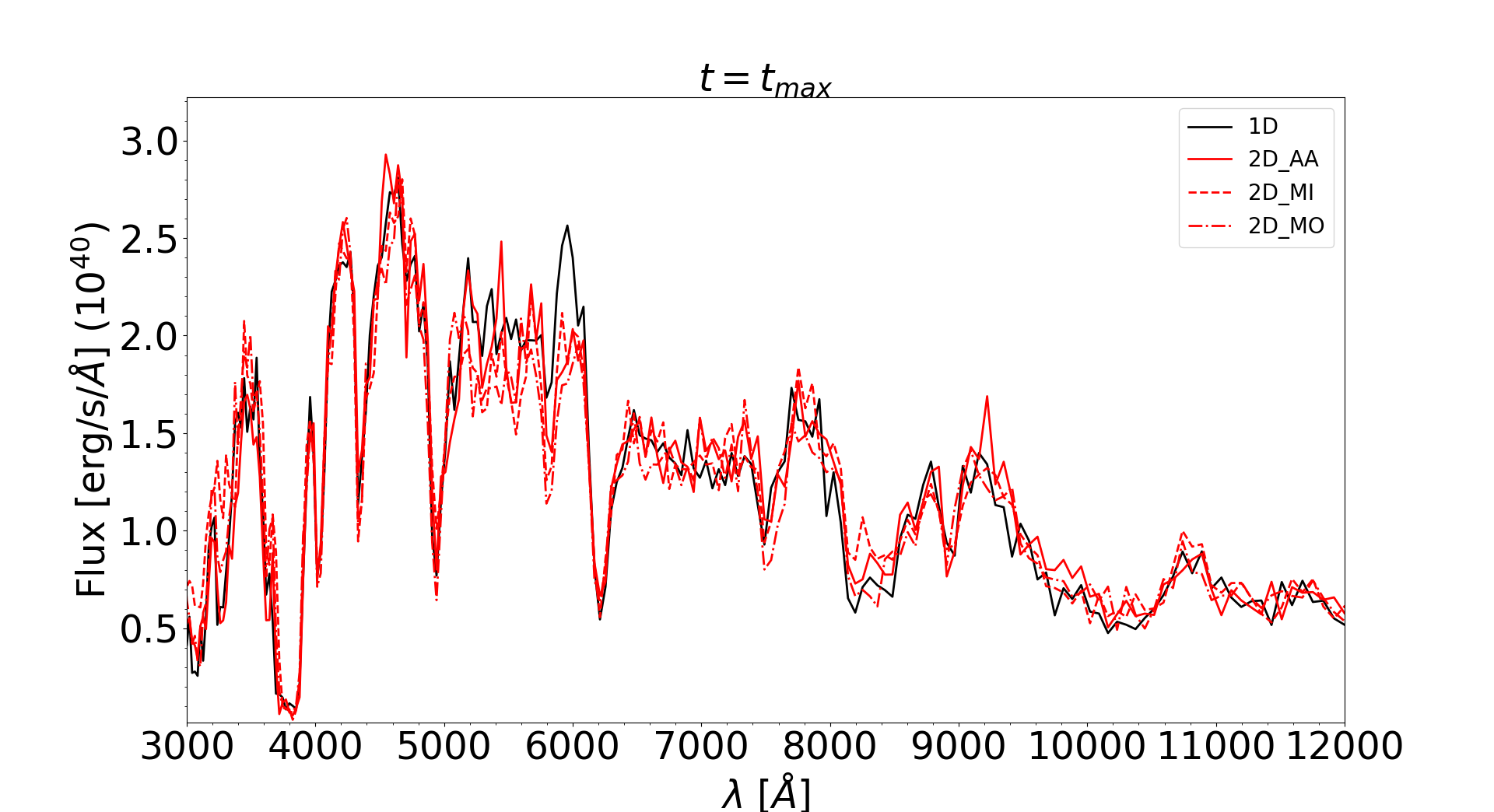}{0.5\textwidth}{}
          \fig{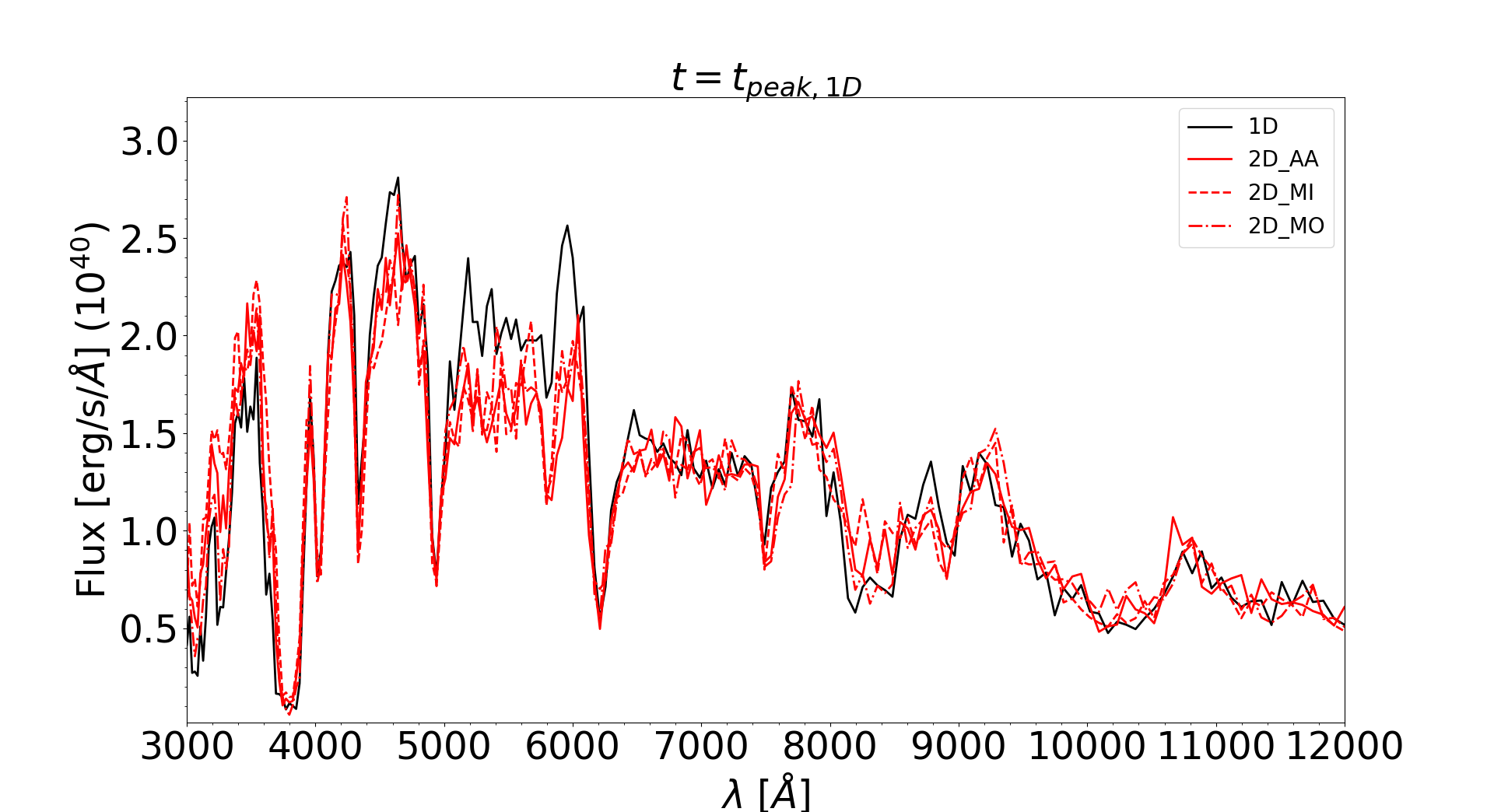}{0.5\textwidth}{}
          }
\gridline{\fig{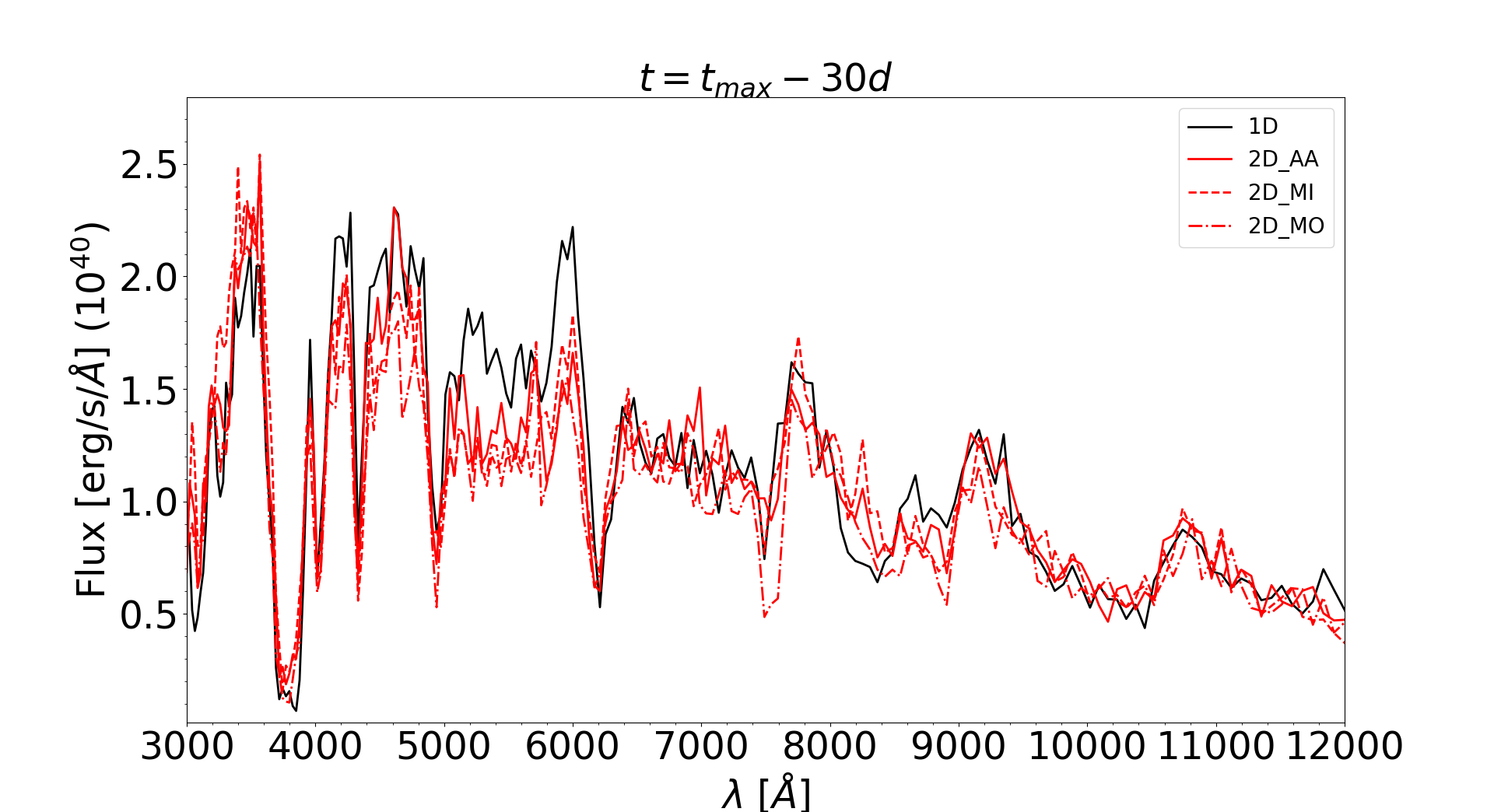}{0.5\textwidth}{}
         \fig{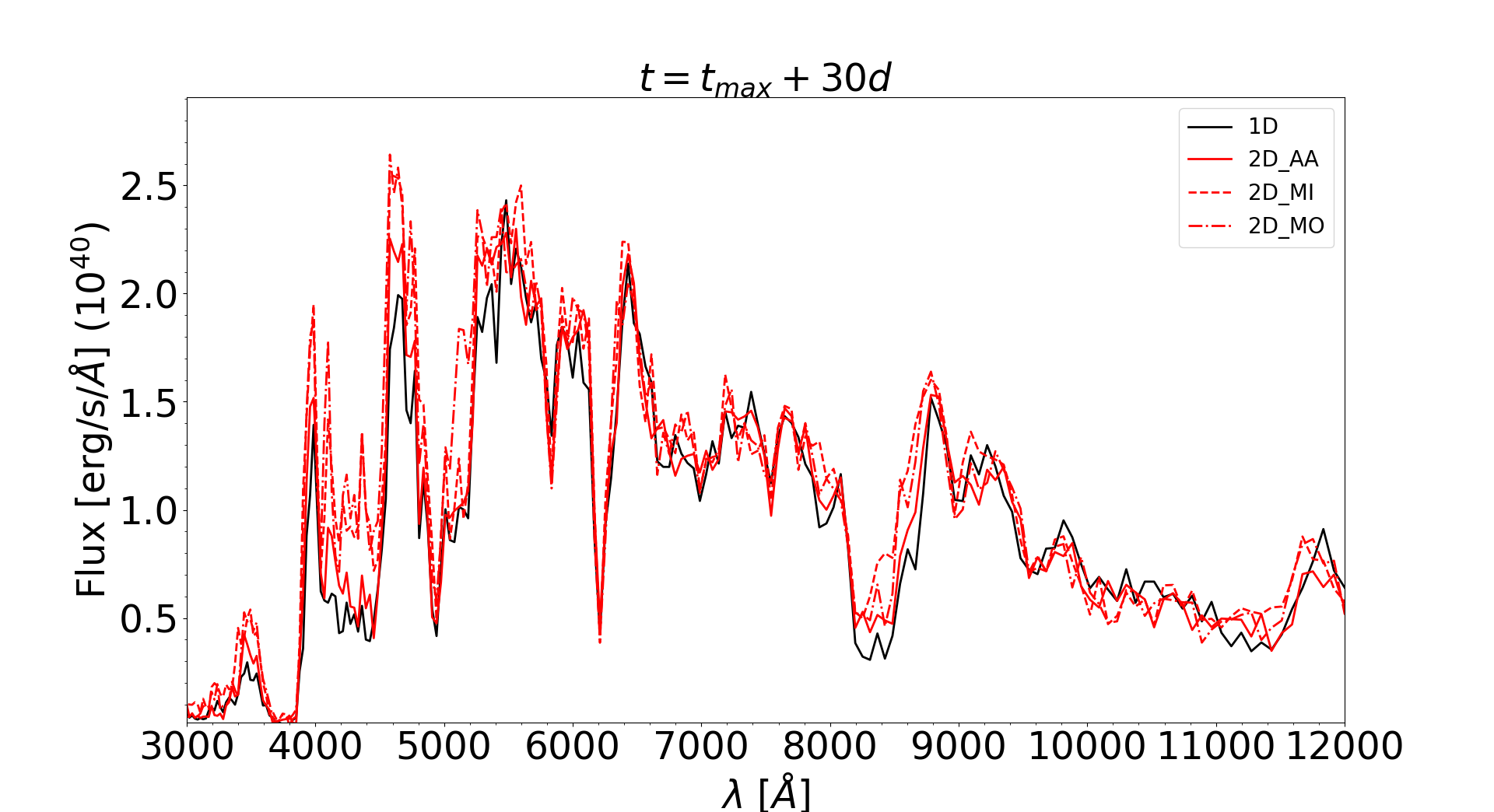}{0.5\textwidth}{}
         }
\gridline{\fig{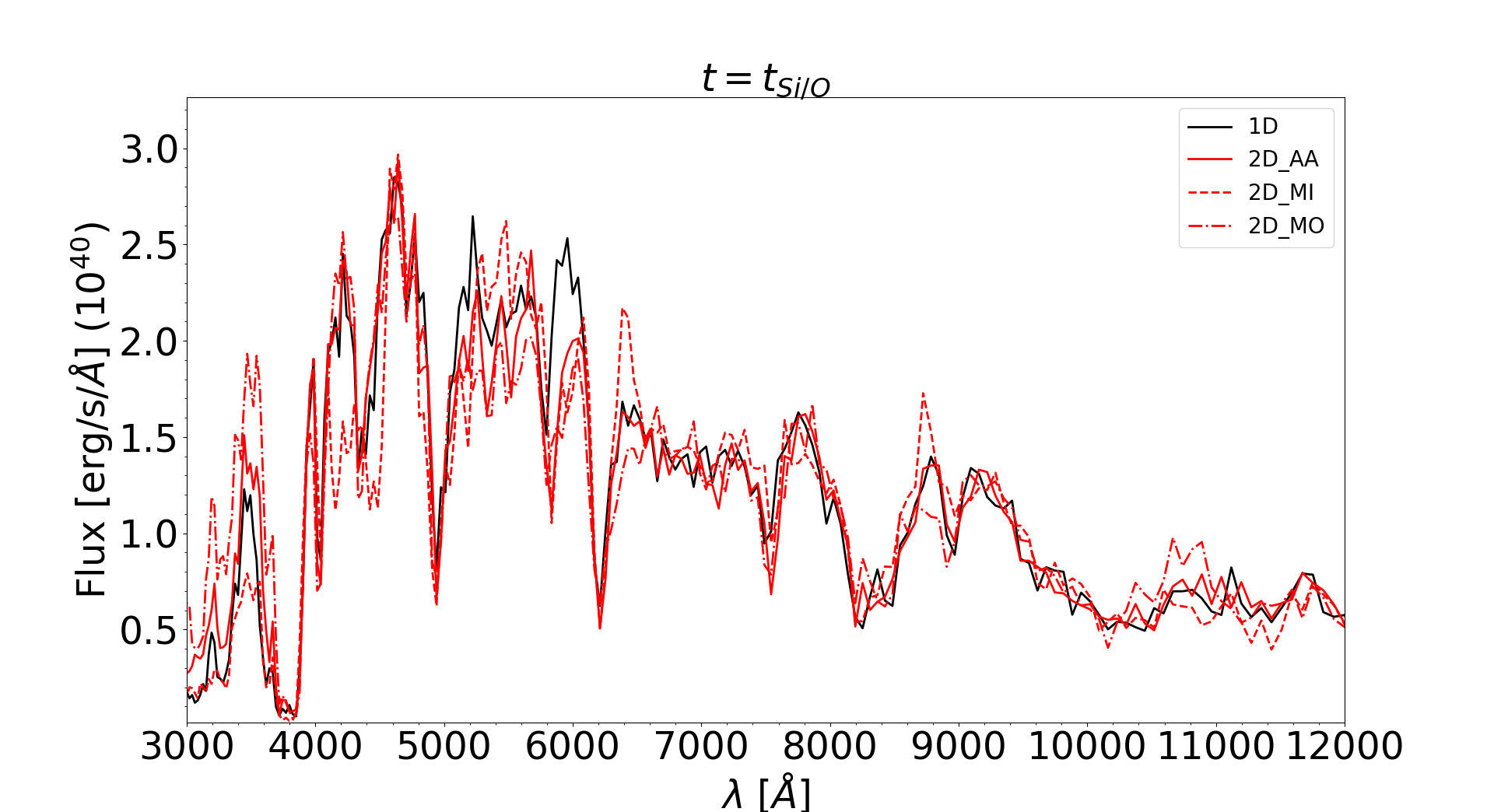}{0.5\textwidth}{}
         \fig{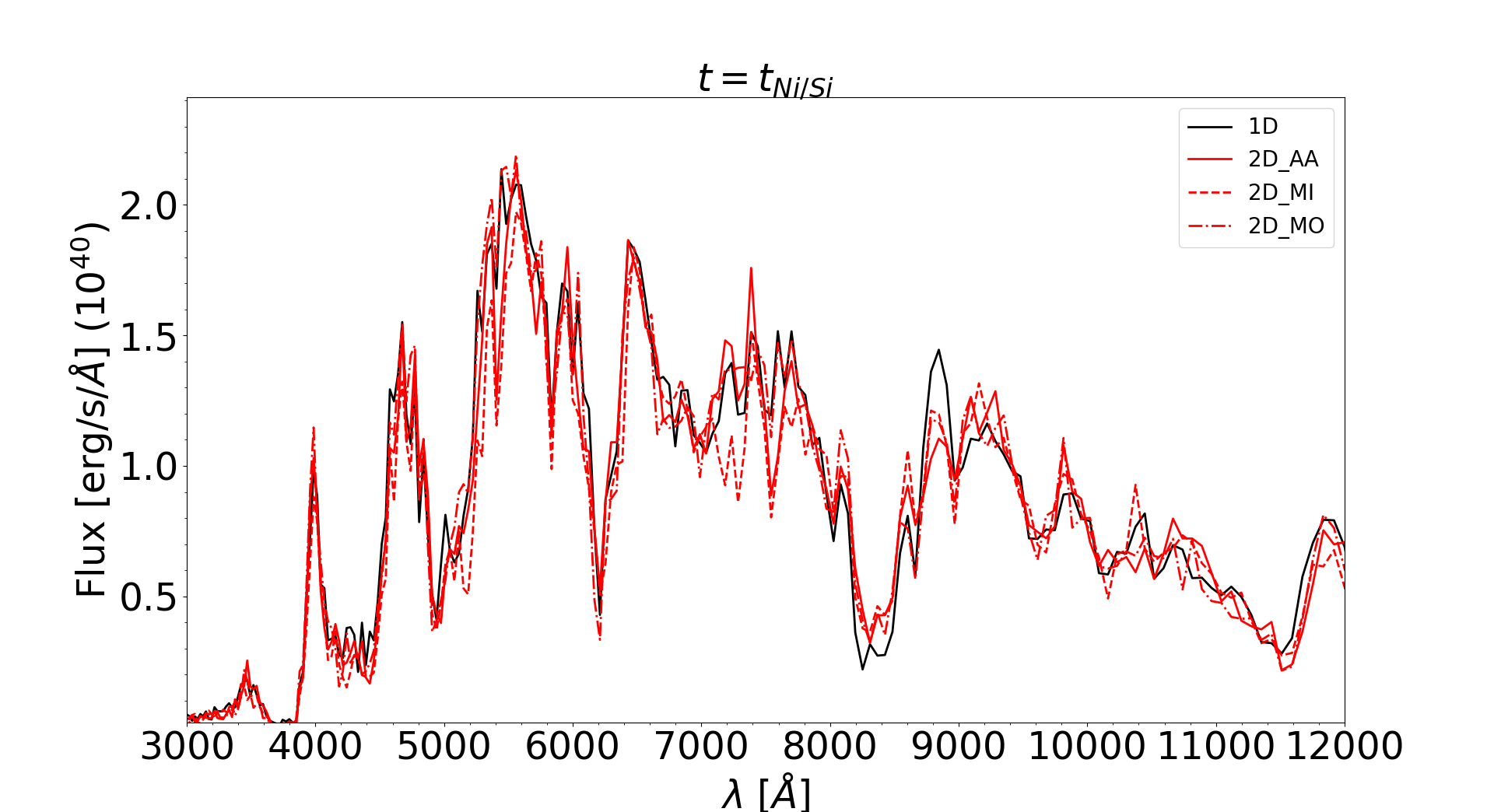}{0.5\textwidth}{}
          }
\caption{Comparisons of synthetic {\it SuperNu} spectra for the {\tt 1D} versus the {\tt 2D\_AA, 2D\_MI, 2D\_MO} models at different epochs: during peak bolometric
luminosity ({\it upper left panel}), at the time of peak bolometric luminosity for the {\tt 1D} model ({\it upper right panel}), at 30~d before and 30~d after peak
bolometric luminosity ({\it middle panels}) and during the transition of the photosphere through regions of high Si/O ({\it lower left panel}) 
and Ni/Si mixing ({\it lower right panel}).
\label{Fig:spec1d2d}}
\end{figure*}

\begin{figure*}
\gridline{\fig{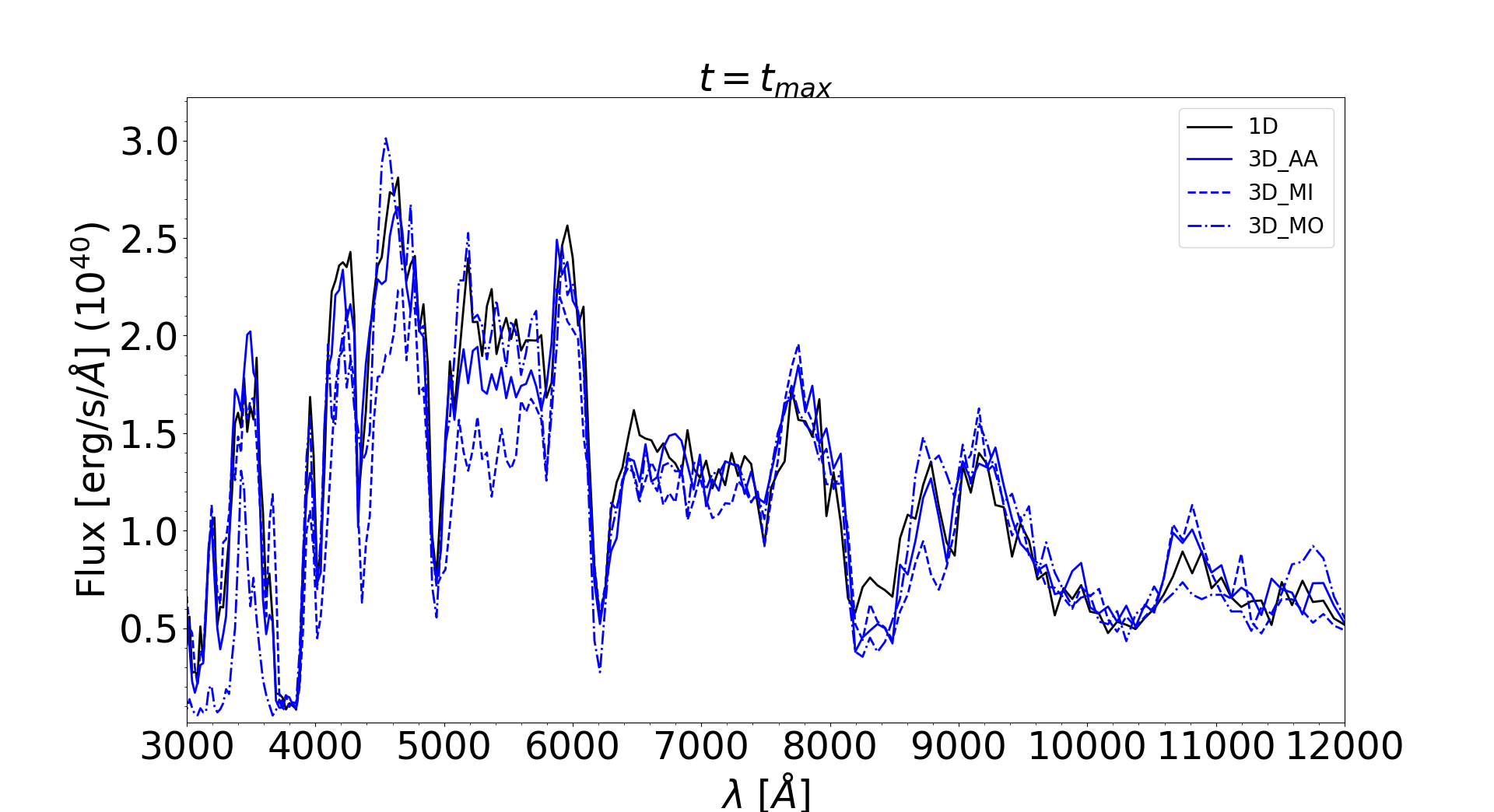}{0.5\textwidth}{}
          \fig{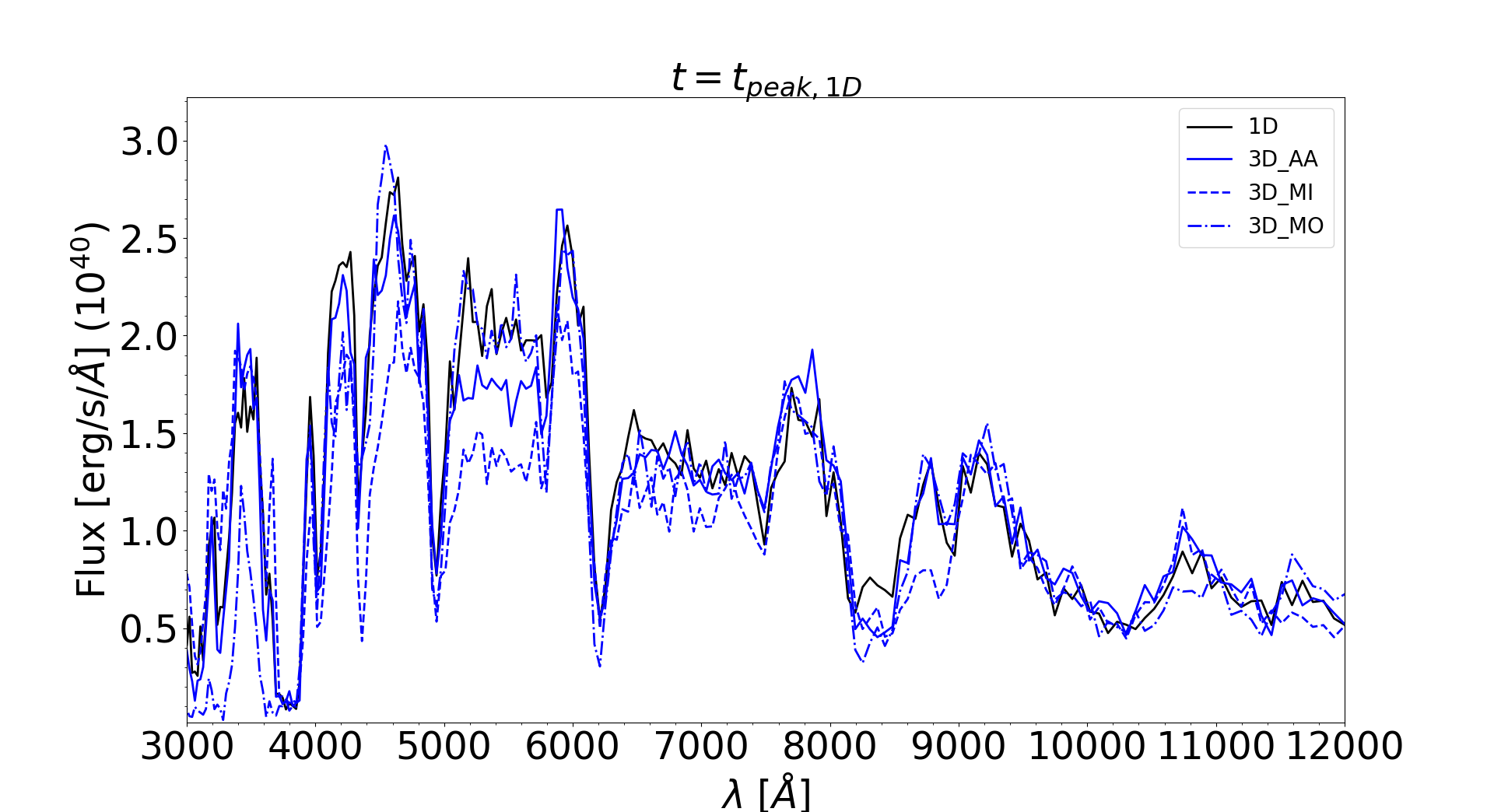}{0.5\textwidth}{}
          }
\gridline{\fig{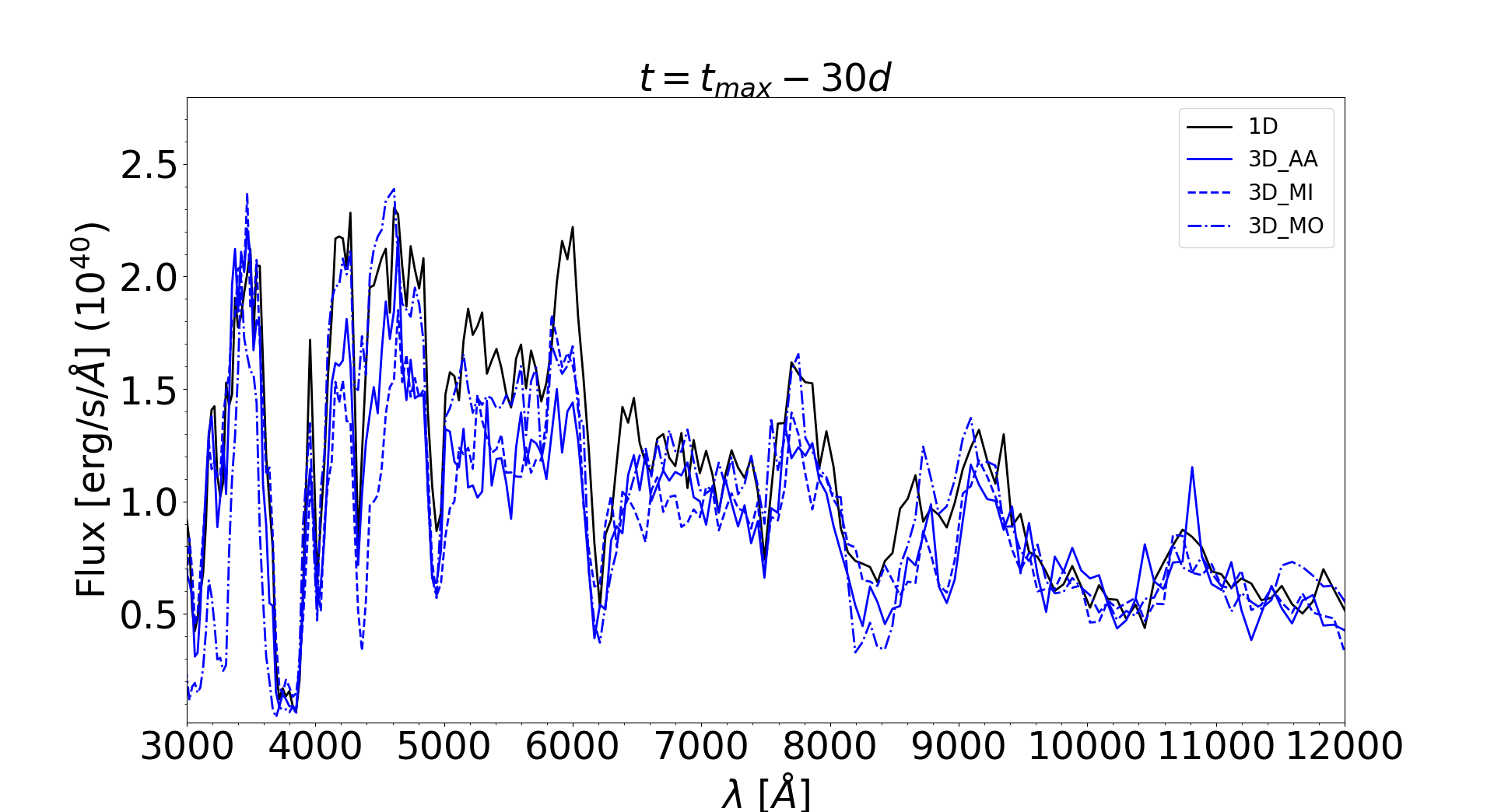}{0.5\textwidth}{}
         \fig{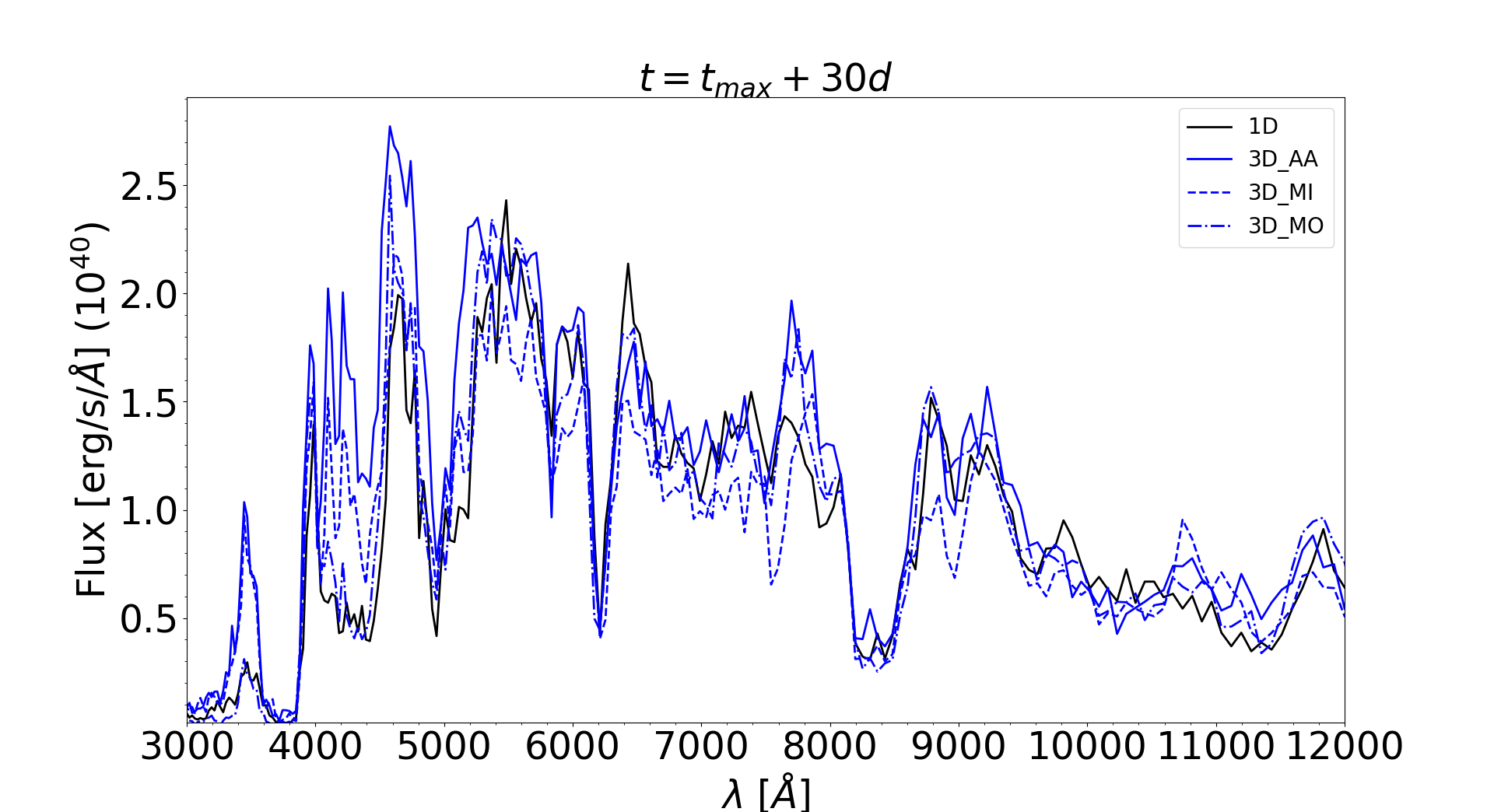}{0.5\textwidth}{}
         }
\gridline{\fig{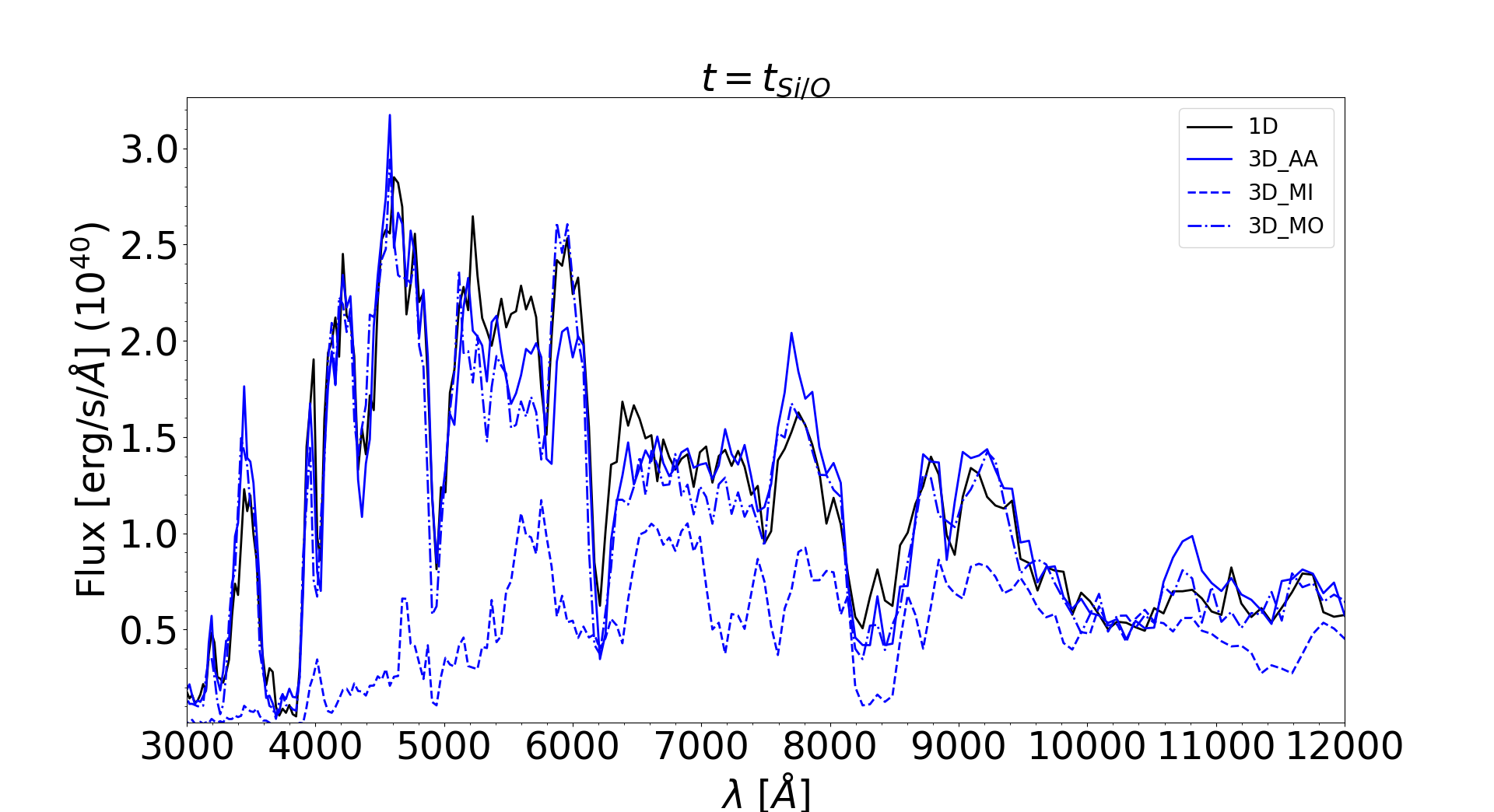}{0.5\textwidth}{}
         \fig{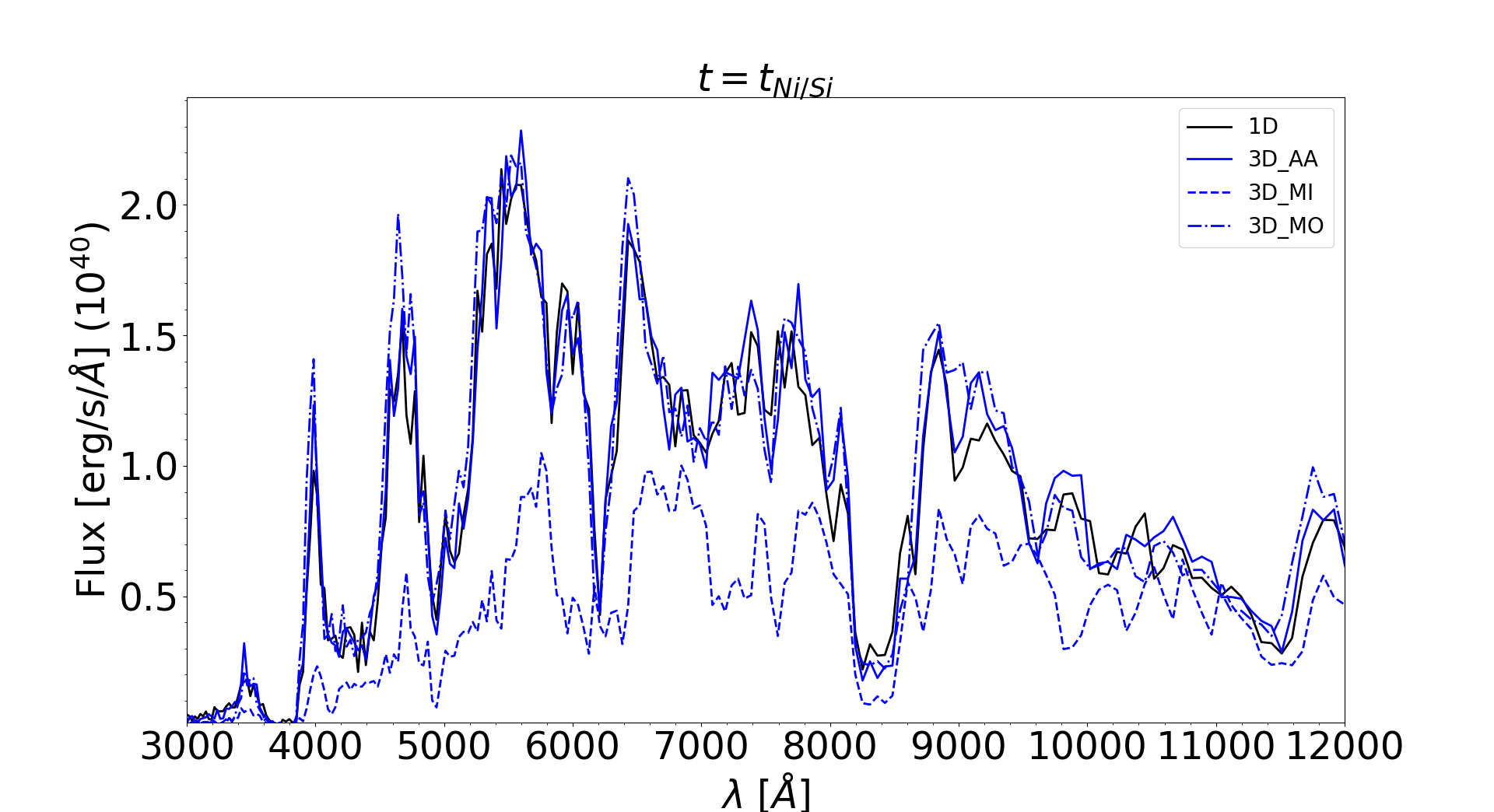}{0.5\textwidth}{}
          }
\caption{Same as Figure~\ref{Fig:spec1d2d} but for the {\tt 1D} versus the {\tt 3D\_AA, 3D\_MI, 3D\_MO} models.
\label{Fig:spec1d3d}}
\end{figure*}

\begin{figure*}
\gridline{\fig{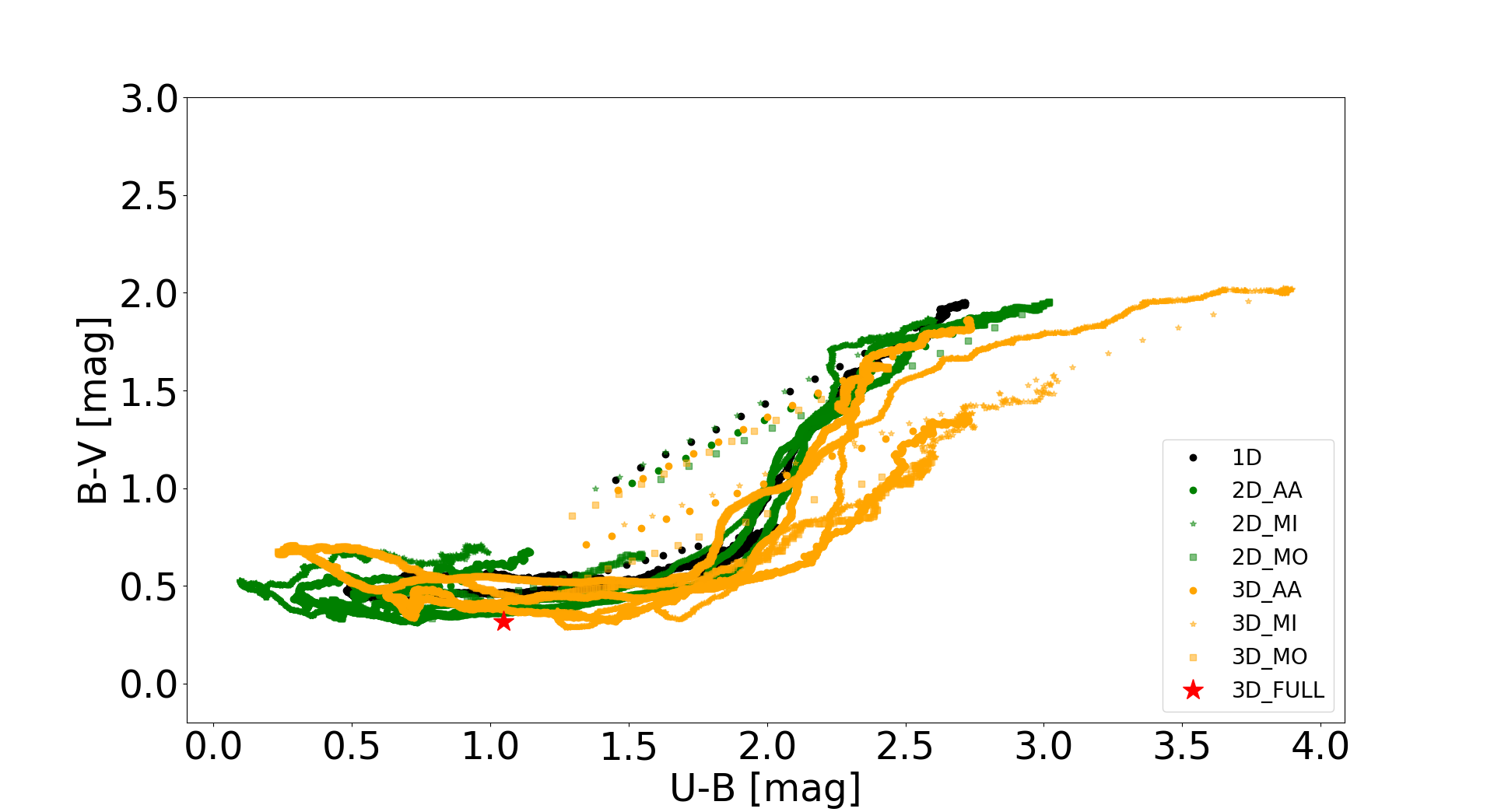}{0.5\textwidth}{}
          \fig{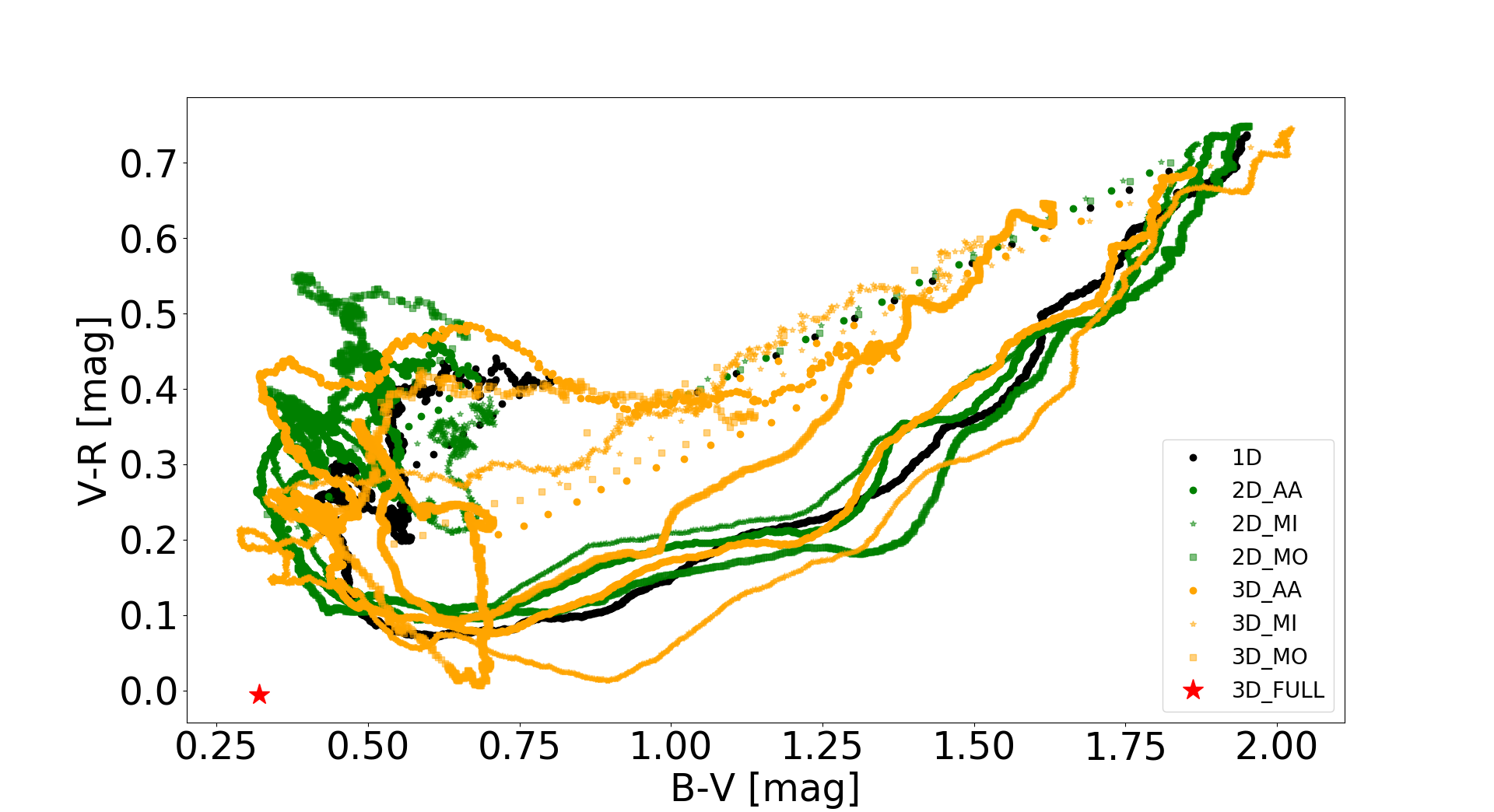}{0.5\textwidth}{}
          }
\gridline{\fig{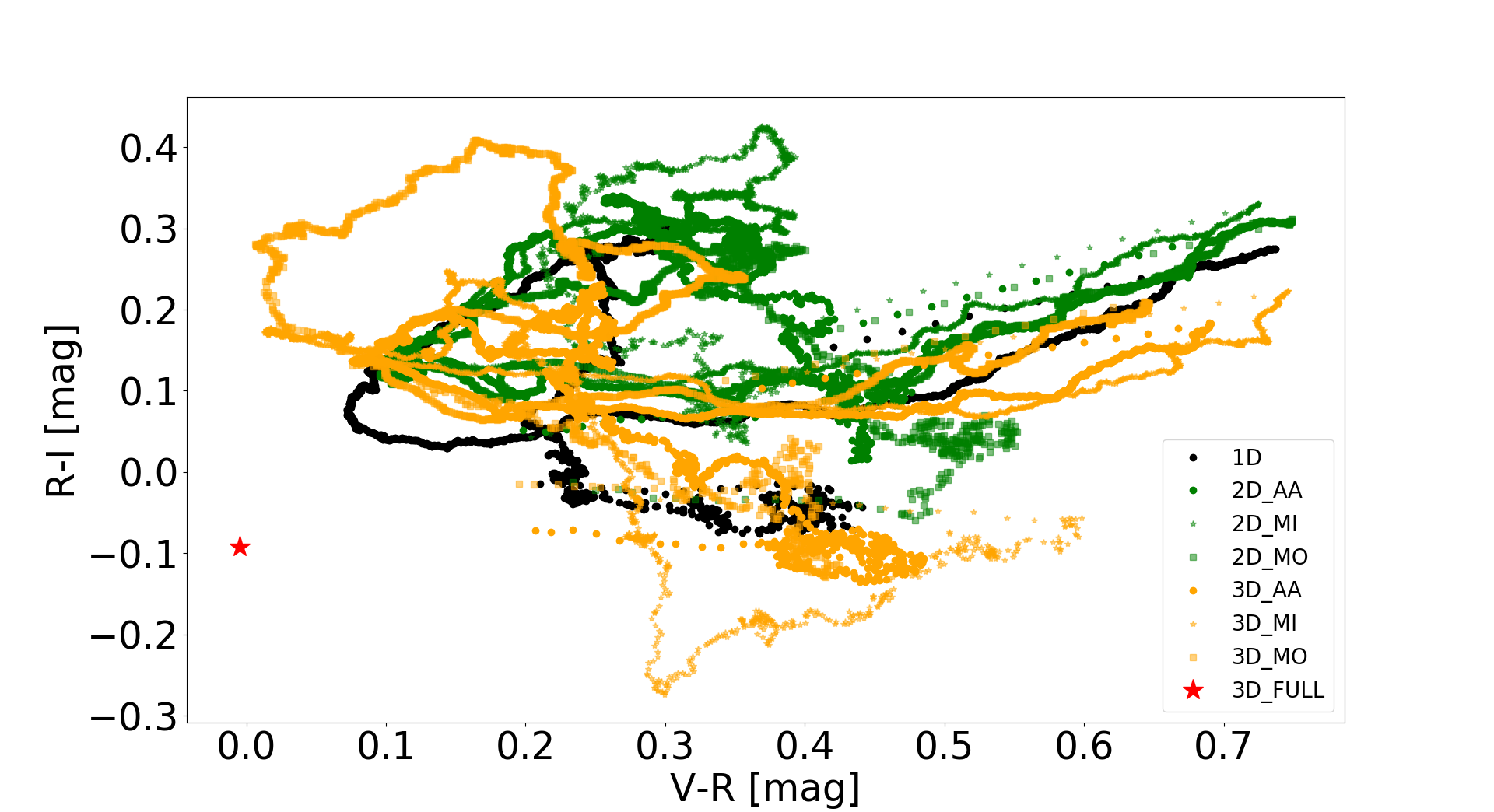}{0.5\textwidth}{}
         \fig{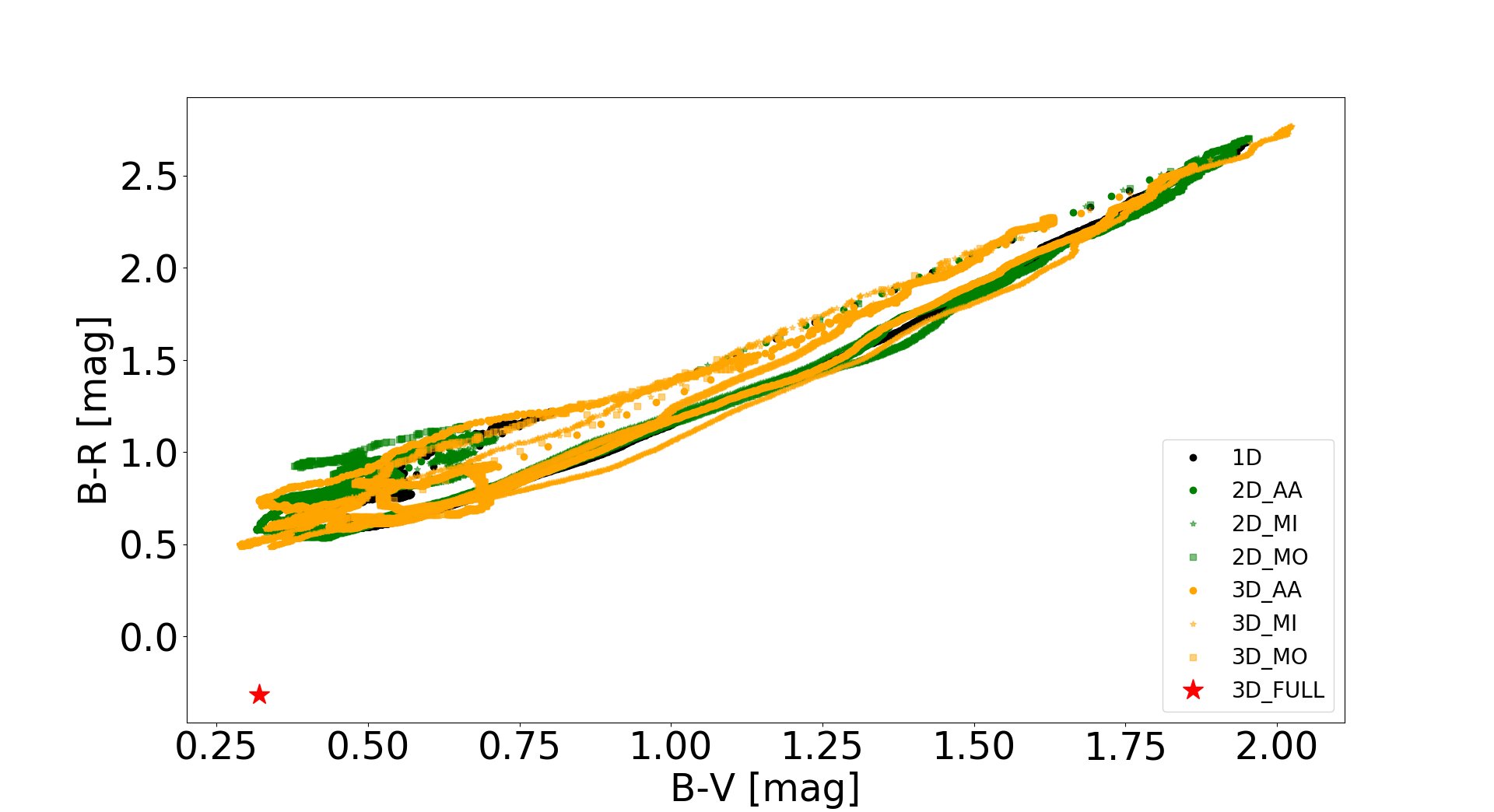}{0.5\textwidth}{}
         }
\caption{Color--Color diagrams for the 1D ({\tt 1D}; black circles), 2D ({\tt 2D\_AA}, {\tt 2D\_MI}, {\tt 2D\_MO}; green circles, star symbols and squares) 
and 3D ({\tt 3D\_AA}, {\tt 3D\_MI}, {\tt 3D\_MO} orange circles, star symbols and squares) models throughout the PISN evolution. The colors for the
{\tt 3D\_FULL} model corresponding to the time of peak luminosity are also shown for comparison (red star symbol).
\label{Fig:colorevol}}
\end{figure*}

In particular, the $\lambda$4130~{\AA}, $\lambda$5051~{\AA} and $\lambda$6355~{\AA} Si~{\rm II} features are very strong. 
The Ca H\&K $\lambda \lambda$3934~{\AA}, 3968~{\AA} doublet is identified in the blue part of the spectrum with blends of iron--peak elements 
populating shorter wavelengths ($\lambda <$~3500~{\AA}). Other prominent absorption features include the Mg~{\rm II} line at $\lambda$4481~{\AA} and the 
O~{\rm I} line at $\lambda$7774~{\AA}. Finally, wavelengths in the range 5100~{\AA}~$<\lambda<$~5800~{\AA} are dominated by Fe~{\rm II} and S~{\rm II}
line blends. The PISN synthetic spectra obtained in this work are therefore consistent with those calculated for hydrogen--poor PISN progenitors in previous
studies indicating spectral evolution that is incompatible with that seen in SLSN--I events at contemporaneous epochs
\citep{2013MNRAS.428.3227D,2015ApJ...799...18C,2019MNRAS.tmp..259M}.

Figures~\ref{Fig:spec1d2d} and~\ref{Fig:spec1d3d} show comparisons of the {\tt 1D} model versus the {\tt 2D\_AA}, {\tt 2D\_MI}, {\tt 2D\_MO} and
{\tt 3D\_AA}, {\tt 3D\_MI}, {\tt 3D\_MO} sets of models, respectively, at similar epochs. In particular, comparisons are shown for synthetic spectra
during peak luminosity ($t = t_{\rm max}$), during peak luminosity as computed in the {\tt 1D} model ($t = t_{\rm max, 1D}=$~159.5~days), 30 days before and 30 days
after peak luminosity and at two phases when the SN photosphere is crossing the Si/O ($t = t_{\rm Si/O}$) and the Ni/Si  ($t = t_{\rm Ni/Si}$) 
interfaces in the PISN ejecta. While the phase corresponding $t_{\rm Si/O}$ is still close to the photospheric phase of the event, the later, $t_{\rm Ni/Si}$
clearly corresponds to the nebular phase of the SN.

The comparison between the 1D and the 2D case indicates that mixing induced by hydrodynamic instabilities shortly after the explosion does
have an effect in the intensity and phase of prominent features in PISN spectra, especially during the earlier phase of the event during the
rise to peak luminosity. In the earlier phase shown in the middle left panel of Figure~\ref{Fig:spec1d2d} ($t = t_{\rm max}-30$~days), the 
O~{\rm I} $\lambda$7774~{\AA} absorption line appears to be stronger for the {\tt 2D\_MO} model as compared to the other cases. Same
holds for the Si~{\rm II} $\lambda$5051~{\AA} feature indicating that outward Si mixing leads to the earlier appearance for these lines
as expected since the photosphere is crossing the mixed Si/O interface sooner as compared to the other cases. In contrast, during
the $t_{\rm Si/O}$ phase (lower left panel of Figure~\ref{Fig:spec1d2d}) the opposite behavior is seen: the {\tt 2D\_MI} model 
Si~{\rm II} $\lambda$6355~{\AA} P Cygni feature is clearly stronger than it is in the other cases for the 2D {\tt P250} profiles.
During the later, nebular phase ($t_{\rm Ni/Si}$) most spectral features are consistent between the different 2D profiles with
the exception of the O~{\rm I}  ($\lambda$7774~{\AA}) absorption line that has become weaker for the {\tt 2D\_MO} case
by that time.

\begin{figure*}
\gridline{\fig{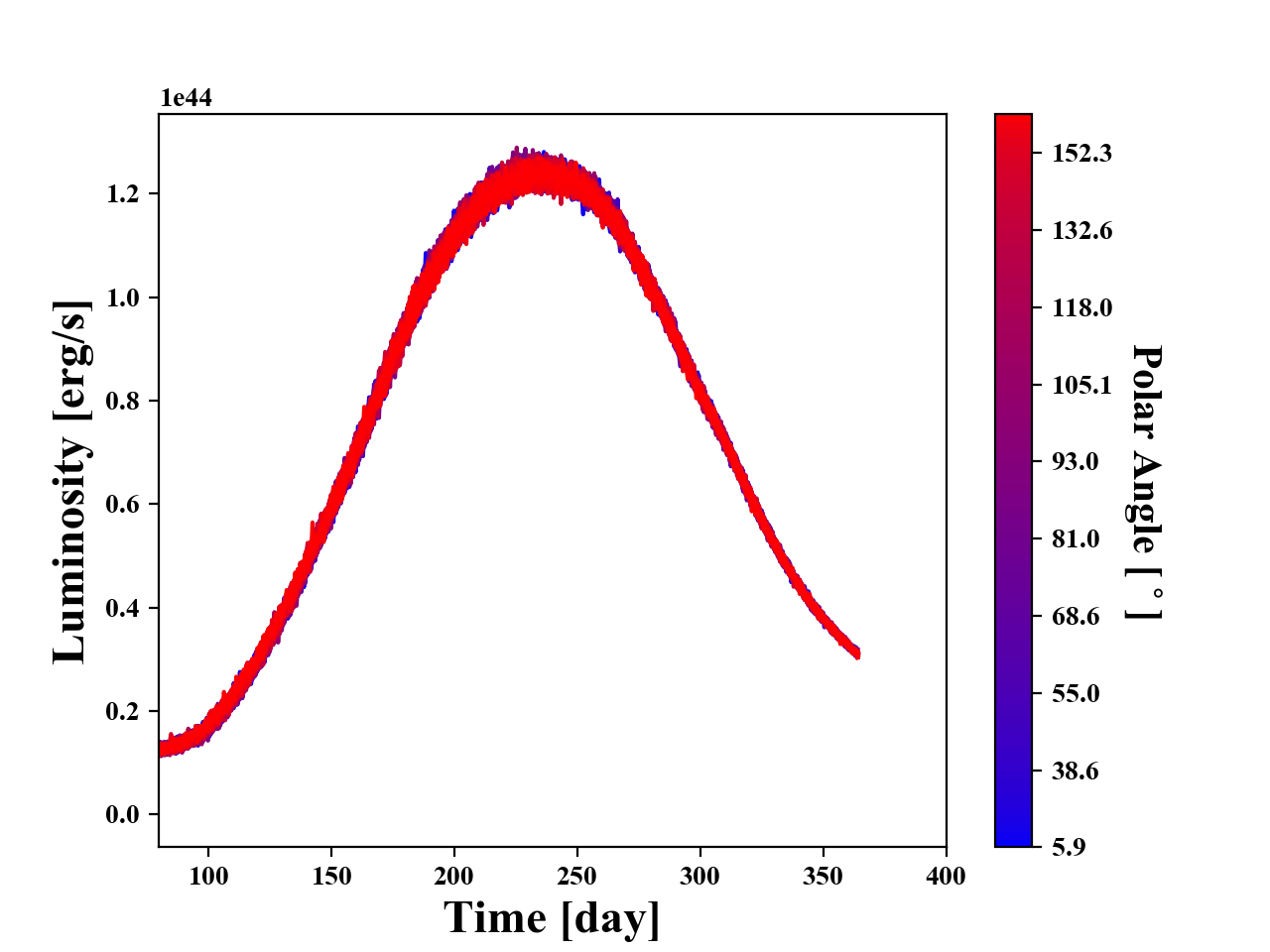}{0.5\textwidth}{}
          \fig{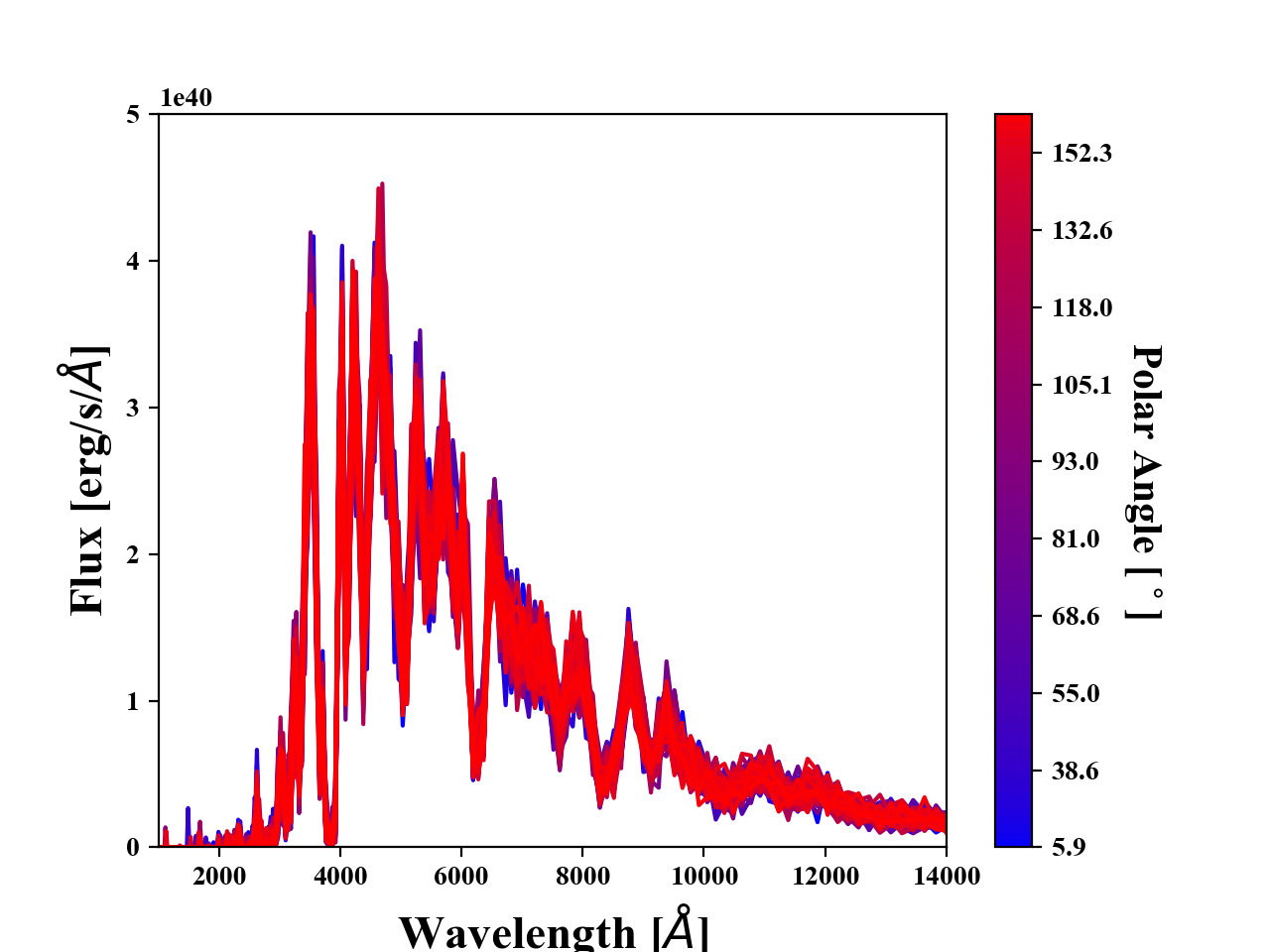}{0.5\textwidth}{}
          }
\caption{Synthetic lightcurves ({\it left panel}) and spectra at the time of peak luminosity ({\it right panel}) as a function of viewing angle from the equatorial plane ($\Omega$) calculated
in the 3D {\it SuperNu} P250 run (model {\tt 3D\_FULL}).
\label{Fig:supernu3D}}
\end{figure*}

\begin{figure}
\begin{center}
\includegraphics[angle=0,width=9cm]{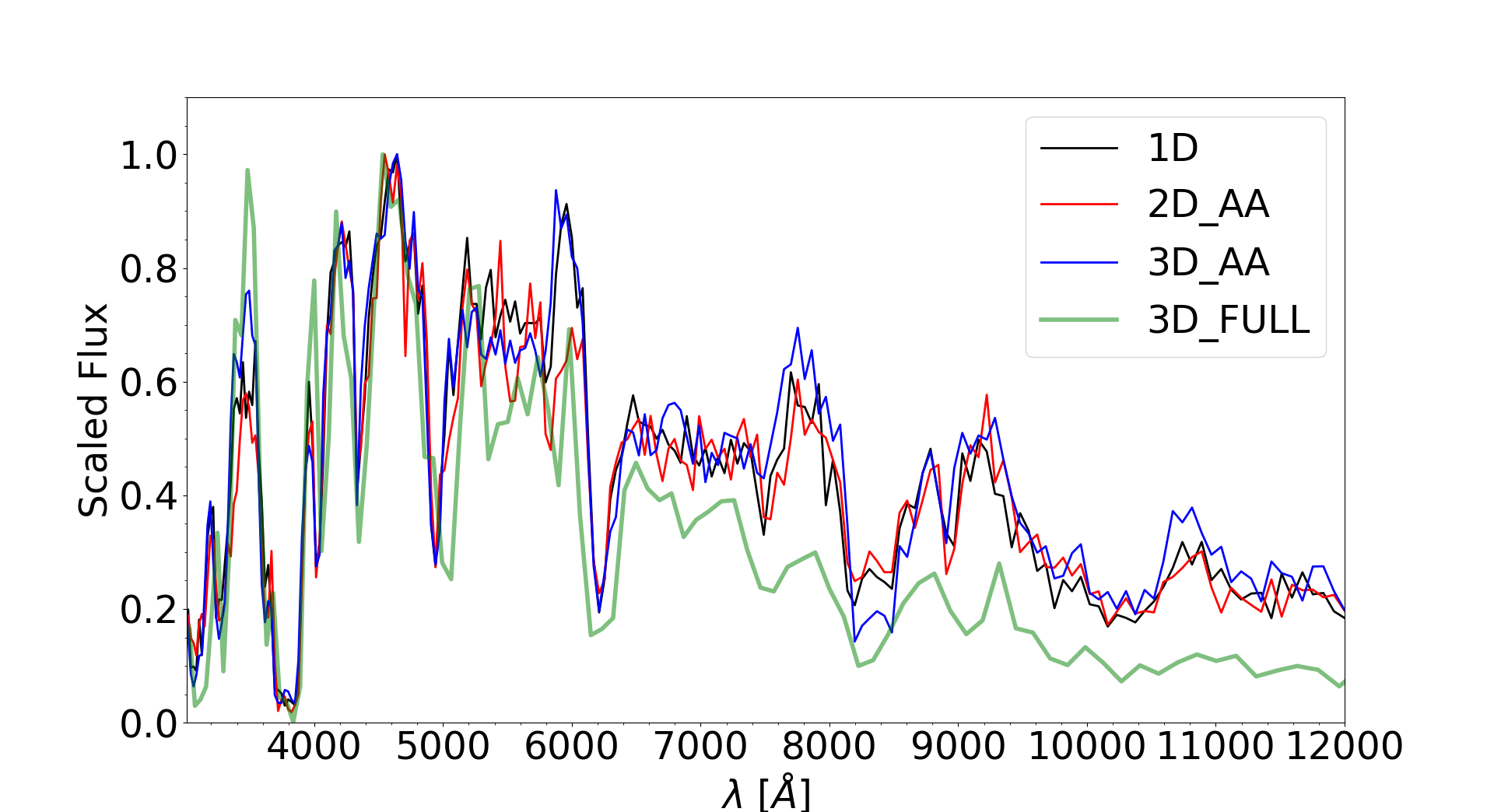}
\caption{Comparison of the {\tt 1D} (black curve), {\tt 2D\_AA} (red curve), {\tt 3D\_AA} (blue curve) and {\tt 3D\_FULL} (thick green curve) {\it SuperNu} synthetic spectra 
at time of peak luminosity ($t = t_{\rm max}$).
\label{Fig:peakspec1d2d3d}}
\end{center}
\end{figure}

The effect of mixing is even more pronounced in 3D than it is in 2D (Figure~\ref{Fig:spec1d3d}). The {\tt 3D\_MI} model possesses a
considerably faster spectroscopic evolution than its {\tt 3D\_AA} and {\tt 3D\_MO} counterparts and is characterized by lower intensity spectral
features especially at later times ($t > t_{\rm max} + 30$~days). As such, during the late phases ($t_{\rm Si/O}$, $t_{\rm Ni/Si}$) the {\tt 3D\_MI} model shows
considerable decline in total radiated flux accross the entire spectrum also reflected in the output bolometric luminosity as discussed
in Section~\ref{snec} (see also Figure~\ref{Fig:lcs_log}). Similarly to the 2D case, the prominent Si and Mg features arise earlier for the {\tt 3D\_MO}
so that they appear stronger as compared to the other two 3D cases at the same phase. Finally, during majority of the SN evolution it appears
that the pronounced mixing present in both the {\tt 3D\_MI} and {\tt 3D\_MO} makes a difference in the total radiated flux as compared to
the angular--averaged case ({\tt 3D\_AA}) where the mixing effects are cancelled out. 
In contrast, the RT mixing in the Ni/Si interface reported by G17 does not appear to have a significant effect in the overall LC or late--time spectra. 
This is in agreement with the findings of \citet{2015MNRAS.454.4357K} who report that extreme -- and hard to realize -- chemical redistribution in the PISN ejecta is 
necessary to see a significant impact on the total radiated luminosity.

Perhaps a better way to illustrate the effects of mixing on the evolution of the radiative properties of PISNe is by plotting color--color diagrams for all
the models investigated here. Figure~\ref{Fig:colorevol} shows four color--color diagrams ($B-V$~vs~$U-B$, $V-R$~vs~$B-V$, $R-I$~vs~$V-R$ and $B-R$~vs~$B-V$)
corresponding to the entire LC evolution of all the models presented in Section~\ref{supernu1D} and to the time of peak luminosity for model {\tt 3D\_FULL} discussed
in the next paragraph. To construct these diagrams, we convolved the Johnson/Cousins ``standard'' filter response curves \citep{Bessell1990,Bessell2012} to the 
computed spectra in {\it SuperNu}. This was done by making use of the {\it Python} {\it speclite 0.8} package\footnote{https://speclite.readthedocs.io/en/latest/filters.html}.
The effect of stronger mixing in 3D is apparent in all the color--color diagrams presented in Figure~\ref{Fig:colorevol}. More specifically, 3D models exhibit redder $V-R$
and $R-I$ colors during the rise to peak luminosity. In addition, there seem to be a correlation between the degree of mixing and the resulting color variance
during the PISN evolution: larger color scatter is observed for the more heavily mixed 3D, followed by somewhat less scatter for the 2D models and even lesser so
for the spherically symmetric 1D models with no mixing. This effect is more discernible in the $V-R$~vs~$B-V$ and $R-I$~vs~$V-R$ color--color diagrams.

The spectroscopic evolution comparisons for models of different degrees of mixing in the PISN ejecta, as captured by multidimensional simulations
of the explosion of a massive H--poor progenitor, clearly illustrates the effects of mixing on the radiative properties of PISNe. Stronger mixing driven
by the RT instability shortly after the explosion leads to faster spectroscopic evolution with key spectroscopic features reaching higher intensities
while occurring at earlier phases. This unveils the potential to decipher the extent of mixing in PISN ejecta using observed time--series of spectra
and potentially study its relation with PISN progenitor structure and explosion energetics.

\subsection{{\it 3D Synthetic Spectra with SuperNu}}\label{supernu3D}

While our 1D {\it SuperNu} models clearly show the impact of mixing on PISN spectra, they are based on lineout profiles along different directions in the 
SN ejecta corresponding to varying degrees in the intesity of RT mixing, specifically between the Si and the O interfaces where it appears to be stronger
(G17). In order to investigate the total effect of 3D mixing on LCs and spectra as a function of viewing angle we designed and ran the first 3D
radiation transport simulation of a PISN in the literature by exploiting the capabilities of {\it SuperNu} (model {\tt 3D\_FULL} in Table 1).

To do so, we mapped the {\it FLASH} 3D PISN profile used for the {\tt 3D\_AA} model into the grid of {\it SuperNu} using 
a script that coarsens from AMR cells to blocks, which are cells in the 3D Cartesian {\it SuperNu} simulation.  
The PISN ejecta is truncated where the velocity drops sharply, so that it can be approximated as homologous.  
The homologous relation between radius and velocity is obtained by a linear regression of the radial projection of the velocity.  
We ensure that the orthogonal component of velocity is small compared to the radial component before simulating with {\it SuperNu}.

Figure~\ref{Fig:supernu3D} shows synthetic LCs and spectra at time of peak luminosity as a function of viewing angle with respect to the pole, $\Omega$. It can be seen that the
radiative properties of the PISN explosion in 3D are virtually independent of viewing angle. This is not surprising provided that the PISN ejecta structure is not
far from spherical symmetry and the effects of large--scale mixing features cancel out. Model {\tt 3D\_FULL} reaches a peak luminosity of $1.3 \times 10^{44}$~erg~s$^{-1}$
that lies in--between the values found in the 1D {\it SuperNu} and {\it SNEC} models (Section~\ref{supernu1D}).
Figure~\ref{Fig:peakspec1d2d3d} presents a comparison of the synthetic spectrum at peak luminosity between the {\tt 3D\_FULL} \
(corresponding to the``edge--on'' vantage point) and the angular--averaged and spherical models. 
The same spectroscopic features are noted, albeit at lower flux levels for the {\tt 3D\_FULL} model. This becomes more apparent for the
O~{\rm I} feature at $\lambda$7774~{\AA} that appears to be much weaker in the full 3D simulation.
Figure~\ref{Fig:colorevol} also shows the color--color diagrams of the {\tt 3D\_FULL} model corresponding to the time of peak luminosity (red star symbol). Interestingly,
the full 3D {\it SuperNu} calculation suggests a bluer color evolution for this PISN model as compared to the 1D mixed profiles. This effect may be attributed to
the more effective leakage of $\gamma$--rays through regions of lower density due to the clumpy structure of the PISN ejecta caused by the growth of hydrodynamic
instabilities, uniquely captured in 3D.

\section{Discussion}\label{disc}

In this paper we explored the effects of hydrodynamic mixing and progenitor model dimensionality on the radiative properties of PISN explosions by
calculating synthetic LCs and time--series of spectra. This was achieved by considering a massive, $M_{\rm ZAMS} =$~250~$M_{\odot}$ (model {\tt P250}), 
progenitor model for which the PISN explosion was simulated in 1D spherical, 2D cylindrical and 3D cartesian geometry
as presented in \citet{2017ApJ...846..100G}. This PISN progenitor was stripped--off its H and He extended envelope prior to reaching the
pair--instability regime and thus exploded as a H--poor star with final mass of $\sim$~127~$M_{\odot}$.

Mixing in the PISN ejecta is triggered mainly due to the growth of the Rayleigh-–Taylor (RT) instability shortly after the collapse of the CO core and
it is found to be stronger in 3D than it is in 2D, as expected from \citet{2014PlPhR..40..451K}. The effects of RT--induced mixing are more pronounced in the Si/O 
interface while minor effects are also observed in the deeper layers such as the Ni/Si interface. 
While it has been shown that unphysically strong outward Ni mixing is required in order to significantly affect the peak luminosities and evolution
timescales of PISN lightcurves, the effect of mixing on the spectra of these events remained unexplored. 

For this reason we post--processed the computed 1D, 2D and 3D model {\tt P250} explosion profiles with the radiation diffusion--equillibrium code
{\it SNEC} and the IMC--DDMC radiation transport code {\it SuperNu} focusing on angles of high inward and high outward mixing present in the Si/O
interface of the PISN ejecta. Furthermore, we run the first full 3D radiation transport simulation of a PISN model with {\it SuperNu}, calculating LCs
and synthetic spectra as a function of viewing angle $\Omega$. Model {\tt P250} is found to produce a slowly--evolving superluminous SN LCs reaching peak
luminosity $> 10^{44}$~erg~s$^{-1}$ and spectra that share a lot of similarities with those of regular Type Ia SNe such as strong features due to intermediate
mass elements (Si, S, Mg and O).

Our calculations reveal that hydrodynamic mixing impacts the spectroscopic evolution of PISNe in a several ways. First, the spectroscopic
evolution for the models where Si mixing is inwards in the Si/O interface is faster as compared to Si ``mixed--outwards'' models and this effect is found to be stronger
in 3D as compared to the 2D case. Consequently, key spectroscopic features reach higher intensities at earlier phases for the Si ``mixed-outwards''
models. Comparisons between the mixed and the angular--averaged 3D profiles show that mixing can also lead to different total radiated flux for
the mixed models, an effect better exemplified by the color evolution of these models. In contrast, the full 3D radiation transport simulation of the {\tt P250} model
indicates that the radiative properties of the explosion are not sensitive to viewing angle because large--scale mixing effects are averaged out. 
This may be different for rapidly rotating PISN progenitors where the overall shape of the expanding SN ejecta becomes highly oblate \citep{2013ApJ...776..129C}.
In addition, the leakage of $\gamma$--rays through regions of lower density captured in 3D leads to bluer color evolution for the PISN model explored here. 
This indicates that multi--color photometry of PISN candidates may hold promise in unveiling the identity of these events in the near future.
Our 3D radiation transport simulation is the first of its kind and showcases the capabilities of {\it SuperNu} to treat explosive outflows with complicated geometries.

The synthetic PISN spectra calculated in this work are in agreement with those presented in previous works and illustrate the difficulty of this 
mechanism to fit the observations of SLSN--I events given the spectroscopic mismatch at similar epochs. Regardless, the capacity of the PISN
model to produce superluminous, slow--evolving transients makes it a relevant candidate for the explosions of rare, extremely massive stars in the local Universe
and for massive primordial Population III stars.
For the latter, cosmological redshift effects may stretch the duration of the observed LC to several years making it hard to distinguish these early stellar explosions
as transient events with upcoming missions such as the {\it JWST} and {\it WFIRST} suggesting that observations of the color evolution and spectra of these events may be
the most suitable way to distinguish them from other sources. Time--dependent, 3D spectroscopic models of these explosions are therefore an important tool
to help identify these elusive cosmic catastrophes in the future.

\acknowledgments

EC would like to thank the Louisiana State University College of Science and the Department of Physics 
\& Astronomy for their support. CF acknowledges support from the US Department of Energy and from Research 
Corporation for Science Advancement. This work used the Extreme Science and Engineering Discovery Environment (XSEDE)  
and the {\it Stampede 2} supercomputer of the Texas Advanced Computing Center ({\it TACC}) at The University of Texas at Austin 
through allocation AST180034. MG, RTW and WPE were supported by the US Department of Energy through the {\it Los Alamos National Laboratory} ({\it LANL}). 
{\it LANL} is operated by Triad National Security, LLC, for the National Nuclear Security Administration 
of the U.S. Department of Energy (Contract No. 89233218CNA000001). The work at {\it LANL} was partially supported by LDRD Grant 20190021DR.

\software
{\tt Matplotlib} \citep{Hunter2007}, {\tt Astropy} \citep{2018AJ....156..123A}.

\bibliography{refs}

\end{document}